\documentclass[fleqn,usenatbib]{mnras}

\usepackage{newtxtext,newtxmath}
\usepackage{footmisc}
\usepackage[T1]{fontenc}

\DeclareRobustCommand{\VAN}[3]{#2}
\let\VANthebibliography\thebibliography
\def\thebibliography{\DeclareRobustCommand{\VAN}[3]{##3}\VANthebibliography}

\usepackage{graphicx}	% Including figure files
\usepackage{amsmath}	% Advanced maths commands
\usepackage{todonotes}
\usepackage{enumitem}
\usepackage{comment}
\usepackage{soul}

\definecolor{lime}{HTML}{A6CE39}
\DeclareRobustCommand{\orcidicon}{%
	\begin{tikzpicture}
	\draw[lime, fill=lime] (0,0) 
	circle [radius=0.16] 
	node[white] {{\fontfamily{qag}\selectfont \tiny ID}};
	\draw[white, fill=white] (-0.0625,0.095) 
	circle [radius=0.007];
	\end{tikzpicture}
	\hspace{-2mm}
}

\foreach \x in {A, ..., Z}{%
	\expandafter\xdef\csname orcid\x\endcsname{\noexpand\href{https://orcid.org/\csname orcidauthor\x\endcsname}{\noexpand\orcidicon}}
}

\def\drm{\mathrm{d}}

\def\kvec{{\bf  k}}
\def\kvecunit{{\bf \hat k}}

\def\Pee{P_{ee}}

\def\qvec{{\bf \tilde {q}}}

\def\exp{{\rm e}}
\def\HI{\ion{H}{I}~}
\def\HII{\ion{H}{II}~}

\newcommand{\ag}[1]{{\color{teal}{[{\bf AG:} #1}]}}

\title[Relating the pkSZ \& the 21 cm power spectra]{Constraining cosmic reionization by combining the kinetic Sunyaev-Zel'dovich and the 21\,cm power spectra}

\author[Georgiev, Gorce, \& Mellema]{
Ivelin Georgiev$^{1}$\thanks{E-mail: ivelin.georgiev@astro.su.se}\orcidA,
Adélie Gorce$^{2, 3}$\orcidB, and Garrelt Mellema$^{1}$\orcidC
\\
$^{1}$The Oskar Klein Centre, Department of Astronomy, Stockholm University, AlbaNova, SE-10691 Stockholm, Sweden\\
$^{2}$Department of Physics and Trottier Space Institute, McGill University, 3600 University Street, Montreal, QC H3A 2T8, Canada\\
$^3$Institut d’Astrophysique Spatiale, CNRS, Université Paris-Saclay, 91405 Orsay, France
}

\date{Accepted XXX. Received YYY; in original form ZZZ}

\pubyear{2023}

\begin{document}
\label{firstpage}
\pagerange{\pageref{firstpage}--\pageref{lastpage}}
\maketitle

% Abstract of the paper
\begin{abstract}
During the Epoch of Reionization (EoR), the ultraviolet radiation from the first stars and galaxies ionised the neutral hydrogen of the intergalactic medium, which can emit radiation through its 21\,cm hyperfine transition. Measuring the 21\,cm power spectrum is a key science goal for the future Square Kilometre Array (SKA), however, observing and interpreting it is a challenging task. Another high-potential probe of the EoR is the patchy kinetic Sunyaev-Zel'dovich effect (pkSZ), observed as a foreground to the  cosmic microwave background temperature anisotropies on small scales. Despite recent promising measurements, placing constraints on reionization from pkSZ observations is a non-trivial task, subject to strong model dependence. We propose to alleviate the difficulties in observing and interpreting the 21\,cm and pkSZ power spectra by combining them. With a simple yet effective parametric model that establishes a formal connection between them, we can jointly fit mock 21\,cm and pkSZ data points. We confirm that these observables provide complementary information on reionization, leading to significantly improved constraints when combined. We demonstrate that with as few as two measurements of the 21\,cm power spectrum with 100\,hours of observations with the SKA, as well as a single $\ell=3000$ pkSZ data point, we can reconstruct the reionization history of the universe and its morphology. We find that the reionization history (morphology) is better constrained with two 21\,cm measurements at different redshifts (scales). Therefore, a combined analysis of the two probes will give access to tighter constraints on cosmic reionization even in the early stages of 21\,cm detections.

\end{abstract}

% Select between one and six entries from the list of approved keywords.
% Don't make up new ones.
\begin{keywords}
observations -- reionization -- cosmological parameters
\end{keywords}

%%%%%%%%%%%%%%%%%%%%%%%%%%%%%%%%%%%%%%%%%%%%%%%%%%

%%%%%%%%%%%%%%%%% BODY OF PAPER %%%%%%%%%%%%%%%%%%

\section{Introduction}
\label{sec:intro}

The epoch of reionization (EoR) is a large-scale phase transition during which the Universe transitioned from a cold and neutral to a hot and ionised state \citep[for a review, see, e.g.,][]{Wise2019}. During the EoR, the first stars and galaxies formed in the densest regions of the Universe due to the accretion of baryonic matter onto dark matter (DM) haloes. The radiation produced by these young stars and galaxies ionised the neutral hydrogen in the intergalactic medium (IGM), forming \HII `bubbles' which progressively grew and overlapped as new ionising sources formed. Because of this, the properties of cosmic reionization contain information on both cosmology and astrophysics.

A powerful probe of the EoR comes from the measurement of the Thomson scattering optical depth $\tau$ from the Cosmic Microwave Background (CMB). As CMB photons emitted during the recombination epoch travel through the intergalactic medium, they scatter off the free electrons produced during reionization. The Thomson scattering generates a polarization signal on
scales larger than the horizon scale during reionization while it suppresses the temperature anisotropies on angular scales lower than the horizon size during reionization. Assuming an instantaneous reionization history, measurements of $\tau =0.051 \pm 0.006$ by the \citet{PlanckCollaborationAghanim_2020} indicate that the midpoint of EoR lies around a redshift of $z_{\mathrm{re}} \sim 8$. On the other hand, observations of the fluctuations of the Lyman-$\alpha$ optical depth caused by the Gunn-Peterson effect in high-$z$ quasar spectra \citep{Bosman2022} and the inferred low mean free path of ionising photons \citep[$\lambda_{\mathrm{MFP}}$,][]{2023Gaikwad} hint that reionization may extend past redshift six and complete by $z_{\mathrm{end}} \sim 5.2$. 

Another promising probe of the reionization process is the kinetic Sunyaev-Zel’dovich (SZ) effect where rest-frame CMB photons scatter off the free electrons along the line-of-sight. As the free electrons from the EoR have a non-zero bulk velocity relative to that of the CMB photons, the latter gain or lose energy, producing secondary temperature anisotropies in the observed CMB \citep{Sunyaev&Zeldovich1980}. This effect occurs during and after the EoR so that it can be divided into two stages. The `homogeneous' kSZ effect is related to the ionised IGM of the post-EoR Universe \citep{ShawRudd_2012}, while the `patchy' kSZ (hereafter, pkSZ) effect has been shown to depend strongly on the morphology of \HII bubbles during reionization \citep{McQuinn2005,Mesinger2012, Alvarez2016,Chen2023}. The pkSZ signal is an integrated observable and the amplitude and peak of its angular power spectrum contains information about, e.g., the duration of reionization and the characteristic sizes of \HII bubbles, respectively \citep{Zahn2005,Iliev2007,GorceIlic_2020}. 
By combining data from the South Pole Telescope\footnote{\url{https://pole.uchicago.edu}} \citep{SPT2004} and the Planck satellite\footnote{\url{www.esa.int/planck}} \citep{PlanckCollaborationAghanim_2020}, \citet{George2015} and, later, \citet{ReichardtPatil_2021} have constrained the amplitude of the pkSZ angular power spectrum to $D^{\mathrm{pkSZ}}_\mathrm{3000} = 3.0 \pm 1.0\,\mu \mathrm{K}^{2}$ using the post-reionization models described in \citet{ShawRudd_2012} for the homogeneous part and \citet{BattagliaTrac_2013} for the patchy component. They deduce a $2 \sigma$ upper limit on duration of reionization from 25 to 75\,\% of $\Delta z <4.1$. However, these upper limits are loosened when accounting for the angular correlation between the cosmic infrared background and thermal SZ power spectrum \citep{ReichardtPatil_2021}. Moreover, the equations relating the pkSZ amplitude to reionization parameters used to derive such constraints are model dependent \citep{Park2013}. \citet{Zahn2012} show the pkSZ amplitude can be suppressed by up to $1.0\,\mu \mathrm{K}^{2}$ due to radiative cooling and depending on the star formation models considered, as they affect the mean gas density within clusters. In addition, in order to properly model the pkSZ power spectrum, the required simulations must be large as well as highly resolved \citep[see, e.g.,][]{ShawRudd_2012}. One way to get around this computational challenge is using a parameterised model calibrated on hydrodynamical simulations \citep[see][for an example]{GorceIlic_2020}.

On the other hand, an extremely promising probe of cosmic reionization comes from the 21\,cm signal emitted by neutral hydrogen within the IGM.
One prospect in detecting this signal is measuring the spherically averaged power spectrum of its spatial fluctuations. This power spectrum contains information about, e.g., the global neutral fraction of the IGM and the growth of the ionising bubbles during reionization \citep{Furlanetto2004, Furlanetto2006, McQuinn2006}. For example, \citet{GeorgievMellema_2021} find a transition scale within the 21\,cm power spectrum which can be directly related to the value of the mean free path of ionising photons $\lambda_{\mathrm{MFP}}$ through an empirical formula $k_{\mathrm{trans}} \approx 2/\lambda_{\mathrm{MFP}}$. 
\begin{comment}
Examples of low-frequency radio interferometers capable of measuring the power spectrum include the Murchison Widefield Array (MWA)\footnote{\url{www.mwatelescope.org}} \citep{Bowman2013}, the Hydrogen Epoch of Reionization Array (HERA)\footnote{\url{https://reionization.org}} \citep{DeBoer2017}, the Low-Frequency Array (LOFAR)\footnote{\url{www.lofar.org}} \citep{vanHaarlem2013}, the Giant Metrewave Radio Telescope (GMRT)\footnote{\url{www.gmrt.ncra.tifr.res.in}} \citep{Swarup1991}, and the forthcoming Square Kilometre Array (SKA)\footnote{\url{www.skatelescope.org}} \citep{2015aska.confE...1K}. A large variety of upper limit values on the 21\,cm power spectrum have been reported at various redshifts and scales by, e.g., LOFAR \citep{Mertens2020}, HERA \citep{HERA2021, TheHERACollaboration_2022}, MWA \citep{Trott2020,Yoshiura2021}, GMRT \citep{Paciga2013}, but no detection has been made and the derived upper limits set only weak constraints on astrophysical quantities \citep{HERA2021, TheHERACollaboration_2022}.
\end{comment}
Examples of low-frequency radio interferometers and the large variety of upper limit values on the 21\,cm power spectrum have been reported at various redshifts and scales by, e.g., the Low-Frequency Array (LOFAR)\footnote{\url{www.lofar.org}} \citep{Mertens2020}, the Hydrogen Epoch of Reionization Array (HERA)\footnote{\url{https://reionization.org}} \citep{HERA2021, TheHERACollaboration_2022}, the Murchison Widefield Array (MWA)\footnote{\url{www.mwatelescope.org}} \citep{Trott2020,Yoshiura2021}, the Giant Metrewave Radio Telescope (GMRT)\footnote{\url{www.gmrt.ncra.tifr.res.in}} \citep{Paciga2013}, as well as the forthcoming Square Kilometre Array (SKA)\footnote{\url{www.skatelescope.org}} \citep{2015aska.confE...1K}. However, no detection has been made and the derived upper limits set only weak constraints on astrophysical quantities \citep{HERA2021, TheHERACollaboration_2022}.

%\ag{Potential for joined analysis: \citet{BeginLiu_2022} for global 21cm signal, \citet{NikolicMesinger_2023} for PS} 
Indeed, the 21\,cm power spectrum is affected by extra-galactic foregrounds from radio bright sources, radio frequency interference and ionospheric activity, which complicate the calibration of the antennas of the radio interferometers and, in turn, its measurement. However, the foreground signal is anticipated to mainly affect the lower-$k$ region of the spectrum in Fourier space and techniques of foreground avoidance, suppression, and subtraction have been considered \citep[see, e.g.,][]{LiuParsons2014,ChapmanAbdalla2013, MertensGhosh2018}.

In this work, we leverage the complementarity of 21\,cm and kSZ observations in probing cosmic reionization in order to obtain significant constraints before a complete measurement of the 21\,cm power spectrum is achieved. In \citet{BeginLiu_2022}, the authors already demonstrate this complementarity at the level of the global 21\,cm signal, showing that a combined analysis makes it possible to reconstruct a model-independent reionisation history, with no assumed parameterisation of the redshift-evolution. Here, we push this analysis a step further by including second-order statistics in the assessment and considering the 21\,cm power spectrum.  %Indeed, the \textbf{longer the reionization process, the larger the} patchy kSZ amplitude, making this observable more sensitive to slowly varying reionization histories. \textbf{In contrast, abrupt reionization histories will be more easily distinguished from smooth foregrounds than slowly varying ones, making the 21\,cm power spectrum de facto more sensitive to the former }\citep{BeginLiu_2022}. %As each probe is either sensitive to the neutral or the ionised medium, a combined analysis leads to a more complete picture of the reionization process. 
We place constraints on cosmic reionization by utilising the fundamental relationship between the power spectra of the 21\,cm and the pkSZ signal from the EoR, which we formalise through a simple yet effective parametric method. %that establishes a connection between the spectra and enables a joint fit of both da`tasets.
%Through an MCMC analysis, we discover that these two observables exhibit complementary characteristics, leading to significantly improved constraints on reionization compared to analysing each data set separately. We present how future measurements of the 21\,cm power spectrum with the SKA and the pkSZ power spectrum measurement are combined to constrain models of cosmic reionization. Our findings demonstrate that a few well-informed measurements of the 21\,cm power spectrum and pkSZ data can precisely determine the reionization history of the Universe.

This paper is organised as follows. In section~\ref{sec:methods}, we describe the physics behind the 21\,cm signal from the EoR, the derivation of the pkSZ angular power spectrum, and the formalism used to formally link the 21\,cm signal and the pkSZ effect, and, in turn, their power spectra. We also introduce the methodology of the Monte-Carlo Markov Chain sampling used in our forecast. Section~\ref{sec:data} briefly outlines the simulation used to check the accuracy of our forecast. In section~\ref{sec:results}, we present our findings for an ideal case, with the detection of both the 21\,cm with SKA-\textit{Low} for two redshifts as well as the pkSZ power spectra. We discuss how the forecast behaves and how constraining it is on the parameters of reionization. In section~\ref{sec:discussion}, we outline the caveats of the forecast and explore special cases. We summarise our conclusion and elaborate on future improvements in section~\ref{sec:conclusions}. Unless stated otherwise, we address comoving megaparsec as Mpc.

\section{Methods}
\label{sec:methods}
In this section, we study the connection between the neutral and ionised density components of the IGM during the EoR by investigating the relation of the 21\,cm and pkSZ power spectra. %We explore how we can utilise the complementary nature between the two probes to better constrain the parameters of reionization. 
We go through the derivation of the 21\,cm power spectrum, where we construct our model based on its relation to the electron auto-correlation power spectrum. Using a parameterisation of the latter, we present how to analytically reconstruct both the 21\,cm and pkSZ signals and generate mock data. Lastly, we describe the methodology of the forecast analysis based on the mock data.

\subsection{Relation between the 21 cm and pkSZ signals and the electron overdensity field}
\label{subsec:2_derivation}

\subsubsection{Derivation of the 21\,cm power spectrum}

The spin-flip transition between the hyperfine states of a neutral hydrogen \HI atom results in the emission or absorption of photons with a 21\,cm wavelength \citep{Field1957}. The abundance of \HI in the IGM makes this radiation an appealing probe of the large-scale structure of the Universe and a scientific goal for radio interferometry telescopes such as the SKA-\textit{Low}. 

The 21\,cm signal from the Epoch of Reionization is seen against the Rayleigh-Jeans tail of the CMB. Assuming small optical depth and neglecting redshift space distortions, the brightness temperature of the signal can be written as
\begin{align}
\label{eq:tb21}
\delta T_{21} (\mathbf{r}, z) = T_{0} (\mathbf{r},z)\, x_{\HI}(\mathbf{r},z) \left[1+\delta_{b}(\mathbf{r},z)\right] \ , 
\end{align}
with \citep{PritchardLoeb_2012}
\begin{align}
T_0 (\mathbf{r},z)& \approx 27 ~\mathrm{mK}\
\bigg(\frac{1+z}{10}\bigg)^{1/2} \bigg(\frac{T_{\mathrm{s}}(\mathbf{r},z) -T_{\mathrm{CMB}}(z) }{T_{\mathrm{s}}(\mathbf{r},z) }\bigg) 
\nonumber
\\ &\times \bigg( \frac{\Omega_{\mathrm{b}}}{0.044}  \frac{h}{0.7} \bigg) \times  \frac{ \Omega_\mathrm{b} h \sqrt{0.15 (1+z)}} {0.023 \sqrt{10 \Omega_{\mathrm{m}}}}.
\end{align}
Here, $\delta_{b}= \rho_b/\bar{\rho}_b -1 $ is the baryonic density fluctuation and the bar denotes a spatial average and $\delta_{\HI}$ is the fluctuation of the neutral fraction $x_{\HI}$ field, i.e. the fraction of the IGM baryons that are neutral. Naturally, $x_{\HI}=1-x_{\HII}$, where $x_{\HII}$ is the ionisation field. The pre-factor $T_{0}$ includes most of the spatially averaged information, including cosmological parameters, and the spin temperature information $T_\mathrm{s}$. We assume the limit where during cosmic reionization the gas in the IGM is sufficiently heated by the first ionising sources so $T_{\mathrm{s}}\gg T_{\mathrm{CMB}}$ and the spin-temperature dependence in $T_0$ is dropped. 

Let us consider the electron fraction field 
\begin{equation}
\label{eq:def_xe}
    x_e \equiv f_\mathrm{H}\, x_{\HII}(1+\delta_b),
\end{equation}
where $f_\mathrm{H}$ is the fractional quantity of electrons per Hydrogen atom. If the first reionization of Helium is considered, $f_\mathrm{H}\simeq1.08$. We distinguish the mass-weighted $x_m$ and volume-weighted $x_v$ average ionised fractions of \ion{H}{I} atoms. By definition, $\langle x_e \rangle \equiv f_\mathrm{H}\, x_m$, and, if we take the fluctuations of each element in equation~\eqref{eq:def_xe}, we have
\begin{equation}
\label{eq:delta_e}
\begin{aligned}
 x_m (1+\delta_e) & = x_v (1+\delta_i)(1+\delta_b),
\end{aligned}
\end{equation}
where we refer to the density and ionisation field perturbations as `\textit{b}' and `\textit{i}', respectively.
Including these definitions in equation~\eqref{eq:tb21} and using the fact that $x_{\HI}=1-x_{\HII}$, we have
\begin{equation}
\begin{aligned}
   \frac{ \delta T_{21}}{T_0(z)} & =(1+\delta_b) - x_m(1+\delta_e)\\
   & = (1+\delta_b) \left[ 1 - x_v (1+\delta_i) \right],
\end{aligned}
\end{equation}
where the spatial- and time-dependence have been omitted for simplicity.
Taking the Fourier transform and averaging, we find
\begin{align}
    \label{eq:recons_21cm_ps_full}
   \frac{P_{21}(k,z)}{T_0(z)^2}  &=  P_{bb}(k,z) +  x_{m}(z)^2 P_{ee}(k,z)  \nonumber \\ &-2x_v(z) \Big( P_{bb}(k,z)+P_{bi}(k,z)+P_{bi,b}(k,z) \Big),
\end{align}
where $P_{21}$ is the power spectrum of the 21\,cm signal fluctuations, $P_{ee}$ of the electron density fluctuations, $P_{ii}$ of the ionisation fluctuations, $P_{bb}$ of the baryon over-density, and their respective cross-terms. 
Note that the cross-terms in the brackets can also be re-expressed as the cross-correlation between the 21\,cm signal and the baryonic density field following \citet{GeorgievMellema_2021}.

To avoid modelling the cross-terms and the three-point correlations, we simplify equation~\eqref{eq:recons_21cm_ps_full} as
\begin{align}
    \label{eq:recons_21cm_ps_simplified}
    \frac{P_{21}(k,z)}{T_0(z)^2} = \left[1-2x_v(z)\right] P_{bb}(k,z) + x_m(z)^2 P_{ee}(k,z).
   % \bar{x}_{i,m}(z)^2 P_{ee}(k,z) =  \frac{P_{21}(k,z)}{T_0(z)^2}  + (2\bar{x}_{i,v}-1) P_{bb}(k,z),
\end{align}
Some further simplifications of equation~\eqref{eq:recons_21cm_ps_simplified} are assumed: We identify the mass-weighted $x_m$ and the volume-weighted $x_v$ ionised fractions and approximate the baryon power spectrum as a biased dark matter power spectrum $P_{bb}(k,z)=b_{\delta b}^2(k) P_{\delta \delta}(k,z)$, such that the fully simplified expression used in this work, unless stated otherwise, is
\begin{align}
    \label{eq:recons_21cm_ps_fit}
    \frac{P_{21}(k,z)}{T_0(z)^2 } = [1-2x_v(z)] \,b_{\delta b}(k){^2} P_{\delta \delta}(k,z) + x_v(z)^2 P_{ee}(k,z).
   % \bar{x}_{i,m}(z)^2 P_{ee}(k,z) =  \frac{P_{21}(k,z)}{T_0(z)^2}  + (2\bar{x}_{i,v}-1) P_{bb}(k,z),
\end{align}
We will investigate the limitations of these assumptions in section~\ref{subsec:3_recons}. Equation~\ref{eq:recons_21cm_ps_fit} yields a relation between the 21\,cm power spectrum and the electron power spectrum $P_{ee}$. We will now look for a similar relation for the kSZ angular power spectrum.

\subsubsection{Derivation of the patchy kSZ angular power spectrum}

The angular power spectrum of the pkSZ effect at multipole $\ell$ can be derived from the electron density power spectrum $\Pee$ under some assumptions\footnote{These assumptions include the Limber approximation, the assumption that the velocity power spectrum is a biased linear matter power spectrum, and the omission of third and fourth order correlation terms \citep{GorceIlic_2020}. The latter implies variations of the order of 10\% ($\sim 0.05 \mu K^{2}$) in the patchy kSZ amplitude \citep[see][for details]{Alvarez2016}}., such that \citep{Mesinger2012,GorceIlic_2020}:
\begin{equation}
\begin{aligned}
\label{eq:def_C_ell_pkSZ}
C_\ell =  & \frac{8 \pi^2}{(2\ell+1)^3} \frac{\sigma_T^2}{c^2} \int \frac{\bar{n}_e(z)^2}{(1+z)^2}\, \Delta_{B,e}^2(\ell/\eta,z)\, \exp^{-2 \tau(z)}\, \eta(z) \, \frac{\drm \eta}{\drm z}\, \drm z ,
\end{aligned}
\end{equation}
with $\sigma_T$ being the Thomson scattering cross-section, $\bar{n}_e(z)$ and $\eta(z)$ are the mean electron density and the comoving distance for redshift $z$, respectively. Moreover, $P_{B,e}$ is the power spectrum of the curl component of the momentum field $\mathbf{q}_{B,e}$ such that $(2\pi)^3 P_{B,e}\, \delta_D(\kvec-\kvec')= \langle \qvec_{B,e}(\kvec)\ \qvec_{B,e}^*(\kvec')\rangle$ where $\delta_D$ is the Dirac delta function, the tilde denotes a Fourier transform, the asterisk a complex conjugate, and we define the dimensionless power spectrum, for a given field $a$ and $b$ at a certain wave number $k$, as $\Delta^{2}_{\mathrm{a,b}}(k)=k^{3} P_{\mathrm{a,b}}(k)/(2\pi^{2})$. We have
\begin{equation}
\label{eq:full_delta_B}
\begin{aligned}
  \frac{\langle \qvec_{B,e}(\kvec)\ \qvec_{B,e}^*(\kvec') \rangle}{(2\pi)^3{\delta_D}(|\kvec - \kvec'|)} \equiv & \frac{2\pi^2}{k^3} \Delta^2_{B,e}(k,z) \\
  = & \frac{1}{(2\pi)^3} \int \drm^3 k'\,  \left[ (1-\mu^2)\, \Pee (|\kvec-\kvec'|)\, P_{vv}(k') \right. \\ & \left. - \frac{(1-\mu^2)\, k'}{|\kvec-\kvec'|}P_{ev}(|\kvec-\kvec'|) \, P_{ev}(k') \, \right],
\end{aligned}
\end{equation}
where where $\mu = \kvecunit \cdot \kvecunit'$ and the $z$-dependencies have been omitted for simplicity. $\Pee(k,z)$ is the power spectrum of the free electrons density fluctuations and $P_{ev}$ is the free electrons density - velocity cross-spectrum. The latter is computed as
\begin{equation}
P_{v e}(k, z) =\frac{f \dot{a}(z)}{k} b_{\delta e}(k, z) P_{\delta \delta}^{\operatorname{lin}}(k, z)
\end{equation}
where $a$ is the scale factor, $f$ the linear growth rate and the bias is defined by the ratio $b_{\delta e}(k, z)^2 \equiv P_{e e}(k, z) / P_{\delta \delta}(k, z)$.

For a given electron power spectrum, both the angular pkSZ and the spherical 21\,cm power spectra can thus be derived and it is this relationship we will explore in this paper. Note that, reciprocally, the pkSZ is effectively an integral of the 21\,cm power spectrum, equation~\eqref{eq:recons_21cm_ps_simplified} can be used to reconstruct the former from the latter, a potential that we investigate in appendix~\ref{app:obs_reconstruction}. 
In section~\ref{subsec:3_fit}, we will use this relationship to perform a joined fit of 21\,cm and pkSZ power spectrum measurements. We now turn to the model used to relate the electron power spectrum to reionization.

\subsubsection{Electron power spectrum}
\label{sec::electron_powe_spectrum}

We assume the $P_{ee}(k,z)$ parameterisation introduced in \citet{GorceIlic_2020}, that is
\begin{equation}
    \label{eq:Pee_model}
\begin{aligned}
P_{e e}(k, z)=&f_{\mathrm{H}}\left[1-x_v(z)\right]  \times \frac{\alpha_0 x_v(z)^{-1 / 5}}{1+[k / \kappa]^3 x_v(z)} \\
& +x_v(z) b_{\delta b}(k, z)^2 P_{\delta \delta}(k, z),
\end{aligned}    
\end{equation}
where $\kappa$ is the electron drop-off frequency, which can be related to the typical size of ionised bubbles during the EoR, log$_{10}(\alpha_{0})$ is the large-scale $P_{ee}$ amplitude, related to the variance of the electron field during reionization. The redshift-independent baryon-dark matter bias $b_{\delta b}(k)$ is given by the adapted \citet{ShawRudd_2012} parameterisation, that is
\begin{equation}
\label{eq::baryon_ps}
b_{\delta b}(k)^2=\frac{1}{2}\left[\mathrm{e}^{-k / k_f}+\frac{1}{1+\left(g k / k_f\right)^2}\right],
\end{equation}
where $k_f = 9.4\,\mathrm{Mpc}^{-1}$ and $g = 0.5$ are constant with redshift and calibrated on the {\sc EMMA} simulation \citep{aubert_2015_EMMA}, which includes coupled radiative transfer and hydrodynamics and is therefore sensitive to the thermal and reionization history.
%We show in section~\ref{app:obs_vs_params} how both the 21\,cm and the pkSZ power spectra vary with the cosmological and reionization parameters used in this model.
The global reionization history $x_v(z)$ in equation~\eqref{eq:Pee_model} is defined according to the following parameterisation \citep{Douspis2015,GorceDouspis_2022}:
\begin{align}
    \label{eq::reion_parametrisation}
    z &< z_{\mathrm{end}}, \quad x_v(z) = 1,\\
    z &\geqslant z_{\mathrm{end}}, \quad x_v(z) = \bigg( \frac{z_{\mathrm{early}}-z}{z_{\mathrm{early}}-z_{\mathrm{end}}} \bigg)^{\alpha},
\end{align}
where $z_{\mathrm{early}} = 20$ is the redshift for which $x_v(z_\mathrm{early}) \approx 10^{-4}$, the ionisation leftover from recombination. Random forests are used to speed up computations and predict the pkSZ power spectrum given the parameter set; see \citet{GorceDouspis_2022} for more details on the training and testing of the random forests. Note that for both kSZ and 21\,cm derivations, the required matter power spectrum is obtained using the Boltzmann integrator \texttt{CAMB}\footnote{Available at \url{https://github.com/cmbant/CAMB}.} \citep{camb1, camb2}. \textbf{} \\

We now have the full framework enabling us to build the spherical 21\,cm and the angular kSZ power spectra given a reionization model, which we will use to jointly fit reionization parameters to mock measurements of these power spectra.

\subsection{Fits to mock data}

\subsubsection{Monte-Carlo Markov chain sampling}

In section~\ref{sec:results}, we perform a joint fit of mock pkSZ and 21\,cm data points, derived as described in section~\ref{subsec:2_derivation}. To do so, we use a version of the \texttt{CosmoMC} Monte-Carlo Markov Chain sampler \citep[MCMC, ][]{LewisBridle2002, Lewis2013}, modified as described in \citet{GorceDouspis_2022}. Rather than sampling the Thomson optical depth $\tau$, we fit for the reionization mid- and endpoint, $z_\mathrm{re}$ and $z_\mathrm{end}$, as well as for the reionization morphology parameters log$_{10}(\alpha_{0})$ and $\kappa$, defined in equation~\eqref{eq:Pee_model}. The model parameters are listed in Table~\ref{tab:rsage_ref_params}. The assumed (flat) prior range is given for sampled parameters and the value of fixed parameters is listed. For each set of model parameters, we generate the corresponding reionization history and electron power spectrum $P_{ee}$. With the latter, we obtain the dimensionless 21\,cm power spectrum $\Delta_{21}^2$ applying equation~\eqref{eq:recons_21cm_ps_simplified}, as well as the pkSZ angular power spectrum according to equation~\eqref{eq:def_C_ell_pkSZ}, at each iteration of the sampler. We evaluate the agreement of our model with (mock) data by assuming two independent Gaussian likelihoods with uncorrelated errors for each data set:
\begin{equation}
    \log \mathcal{L}_\mathrm{tot} = -\frac{(\Delta_{21,\mathrm{true}}^2 - \Delta_{21,\mathrm{model}}^2)^2}{2 (\Delta^2_\mathrm{noise} + \Delta^2_\mathrm{var})} - \frac{(C_{\ell,\mathrm{true}}^\mathrm{pkSZ}-C_{\ell,\mathrm{model}}^\mathrm{pkSZ})^2}{2 \sigma^2_\mathrm{pkSZ}} ,
\end{equation}
where uncertainty in the measurements $\Delta^2_\mathrm{noise}$, $\Delta^2_\mathrm{var}$, and $\sigma_\mathrm{pkSZ}$ are calculated in section~\ref{subsec:2_errors}. The true model, used to obtain the mock data points, is generated from equations~\eqref{eq:recons_21cm_ps_simplified} and~\eqref{eq:def_C_ell_pkSZ} for the parameters in Table~\ref{tab:rsage_ref_params}.

We employ the Gelman-Rubin test \citep{GelmanRubin1992} to asses the convergence of the MCMC. The $R$ parameter used in this method represents the variance of the chain means compared to the mean of chain variances. In the cases where $R \lesssim 10^{-2}$, we consider the chains to have converged. All parameter values in this work are reported as the marginalised posterior probability's maximum, which is more suitable for skewed distributions. Confidence intervals correspond to intervals with the highest probability density at 68 \%. 

\begin{table}
    \centering
    \begin{tabular}{lcc}
        Parameter & Ref. value & Prior \\
        \hline\hline
        $h$ & 0.681 & N.A. \\
        $\Omega_b $ &  0.050 & N.A. \\ 
        $\Omega_m$ & 0.302 & N.A. \\
        $n_s$ & 0.96 & N.A. \\
        $A_s$ & $2.10\times 10^{-9}$ & N.A. \\
        $z_\mathrm{re}$ & 7.37 & [5.0, 10.0] \\
        $z_\mathrm{end}$ & 6.15 & [4.5, 9.] \\
        d$z = z_{\mathrm{re}} -z_{\mathrm{end}}$ & 1.22 & [0.5, 5.5] \\
        $\log[\alpha_0/\mathrm{Mpc}^3]$ & 3.12 &  [2.5, 4.5] \\
        $\kappa / \mathrm{Mpc}^{-1}$ & 0.145 & [0.05, 0.25] \\
        %$\tau$ & 0.065 & N.A. \\
        \hline 
        
    \end{tabular}
    \caption{Reference cosmological and reionization parameters used to generate the mock data, based on the tuned to best fit the \texttt{RSAGE} simulation. All parameters apart from $z_\mathrm{re}$, $z_\mathrm{end}$,  log$_{10}(\alpha_{0})$, and $\kappa$ are fixed.}
    \label{tab:rsage_ref_params}
\end{table}

\subsubsection{Error estimation}
\label{subsec:2_errors}

Following equation~(11) from \citet{Mellema2013}, also used in \citet{2015aska.confE...1K}, we estimate the dimensionless noise power spectrum for different experiments and observation strategies with
\begin{equation}
\label{eq:tel_noise}
    \Delta^{2}_{\mathrm{noise}}(k, z) \equiv k^{3/2} \lambda_{21}(z) \times \frac{2}{\pi}  \sqrt{\frac{D^{2}_{c}(z)\, \Delta D_{c} }{A_{\mathrm{eff}}}} \frac{T_{\mathrm{sys}}(z)^2}{B\, t_{\mathrm{int}}}  \frac{A_{\mathrm{core}}A_{\mathrm{eff}}}{A_{\mathrm{coll}}^{2}}  ,
\end{equation}
where $\lambda_{21}(z)$ is the redshifted 21\,cm wavelength, $B$ is the bandwidth centred on redshift $z$ and $\Delta D_c$ is its length in comoving Mpc. We write the system noise $T_{\mathrm{sys}} = 100 + 300  (\nu_{\mathrm{obs}} / \nu_{0})^{-2.55}$ K, for $\nu_{0} = 150$ MHz and $\nu_{\mathrm{obs}}(z)=c/\lambda_{21}(z)$ is the observed 21\,cm frequency. The total integration time is $t_{\mathrm{int}}$, $D_{c}(z)$ is the comoving distance to redshift $z$, and $n_{\mathrm{base}}$ the number of baselines, roughly equal to the number of antennas squared $N_a^2$. The total collecting area of the telescope is $A_{\mathrm{coll}}$, such that $A_{\mathrm{coll}} \equiv N_a \pi R_a^2$ where $R_a$ is the radius of the antenna. We have $A_{\mathrm{core}}$ the core area of the array and $A_{\mathrm{eff}}$ the effective collecting area of each antenna. We use the values presented in Table~\ref{tab:noise_est} to estimate the noise for four different experiments: MWA, HERA, LOFAR, and SKA-\textit{Low}. Note that for the latter and LOFAR, we consider each station as a single antenna. 

\begin{table}
    \centering
    \begin{tabular}{lcccc}
        Parameters & SKA-\textit{Low} & MWA & LOFAR & HERA \\
        \hline\hline
        $B$, MHz & 10  & 10& 10  & 10 \\
        $N_{\mathrm{ant}}$ &  224 & 128  &  12& 350 \\ 
        $A_{\mathrm{eff}}$, m$^{2}$ & 600 & 21.5 & 804 & 150 \\
        $R_{\mathrm{core}}$, m  & 500 & 150 & 150 & 150  \\
        $\theta$, deg$^{2}$  & 3$^{2}$ & 24.7$^{2}$ & 3.7$^{2}$ & 8$^{2}$  \\
        \hline 
    \end{tabular}
    \caption{Parameters describing the mock observations by different telescopes used to derive our 21\,cm power spectrum errors (equation~\ref{eq:tel_noise}).}
    \label{tab:noise_est}
\end{table}

We also include the contribution of sample variance to the uncertainty of the measurement following the relation \citep[see eqs.~9 \&10 of][]{Mellema2013}:
\begin{equation}
\label{eq::cosmic_variance}
    \Delta^{2}_{\mathrm{ var}}(k, z) = 0.01 \, \bigg(\frac{k}{0.1 \, \mathrm{Mpc}^{-1}}\bigg)^{-3/2} \bigg( {\frac{V}{1 \,\mathrm{cGpc}^{3}}}\bigg)^{-1/2} \bigg( {\frac{\epsilon}{0.5}}\bigg)^{-1/2} \Delta^{2}_{21},
\end{equation}
where $\epsilon$ is the  logarithmic binning of $\Delta k = \epsilon k$ and $V$ is the survey size, which can be expressed as
\begin{equation}
    V = 0.1\, \mathrm{Gpc}^{3} \times \bigg( \frac{\theta}{5^{\circ}} \bigg)^{2} \bigg( \frac{B}{12\, \mathrm{MHz}} \bigg) \,\big[(1+z)^{1/2}-2 \big],    
\end{equation}
and $\theta$ is the field of view.

\begin{figure}
    \centering
    \includegraphics[width=\columnwidth]{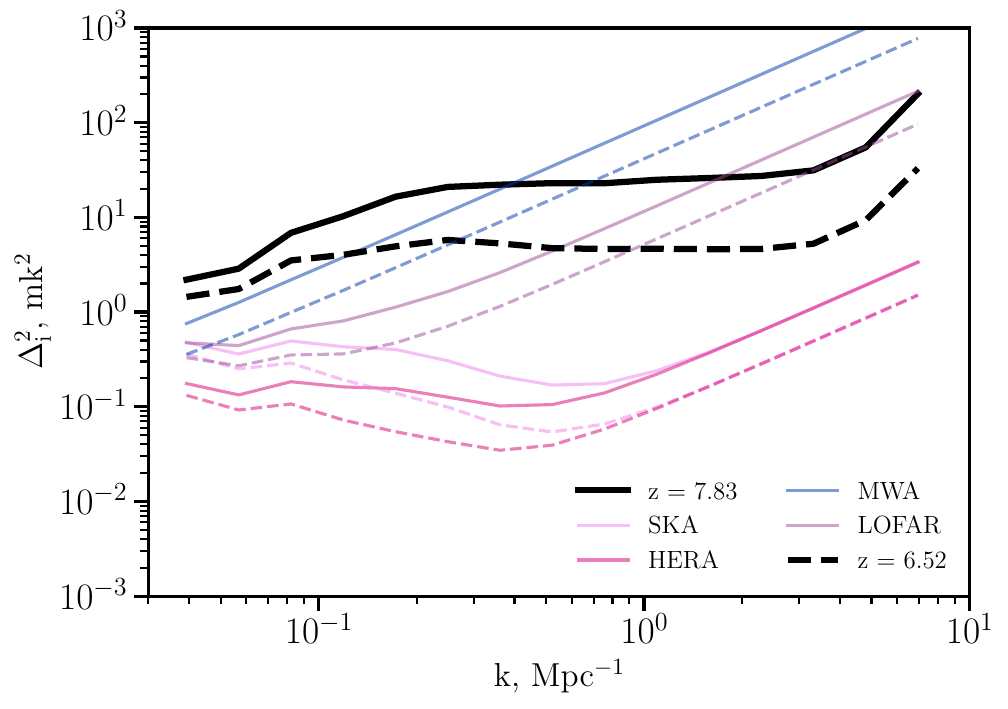}
    \caption{Measurement errors on the 21\,cm power spectrum from {\sc RSAGE} (seen in black) for 1000~hours of integration with different interferometers, corresponding to different colours. These include noise and sample variance errors. The noise and sample variance errors across different $k$-scales are presented for redshifts $z = 6.5, 7.8$ as thick solid and dashed lines, respectively.}% Note that the behaviour of the noise and amplitude is consistent by that reported by \citet{Raste2021} albeit slightly more conservative.}, which can be attributed to the differences in the methodology.}
    \label{fig:tel_noise}
\end{figure}

An example of measurement errors for an integration time of 1000\,hours is presented in Fig.~\ref{fig:tel_noise} for each radio telescope considered for two redshifts, chosen following the upper limits of \citet{Trott2020}. The 21\,cm power spectrum for each redshift is also included. Note that the sample variance dominates the total error budget for the SKA-\textit{Low} at $k \lessapprox 0.3$\,Mpc$^{-1}$ because of its strong $k$-dependency ($\Delta^{2}_{\mathrm{var}} \propto k^{-3/2}$). The opposite is true of the noise estimate for MWA, which has a larger field of view and as $\Delta^{2}_{\mathrm{var}} \propto 1/ \theta$ has a lower error due to sample variance but a higher uncertainty on the noise due to its configuration.% The noise power spectrum scales with $k^{3/2}$ and decreases with redshift.

Regarding the measurement errors on the pkSZ angular power spectrum, we choose a ten per cent uncertainty on the data point throughout this work. We base this on the current constraint of the total kSZ amplitude $D^{\mathrm{pkSZ}}_\mathrm{3000} = 3.0 \pm 1.0\,\mu \mathrm{K}^{2}$ \citep{ReichardtPatil_2021}, assuming that in future observations longer observation times will allow to reduce the noise, whilst numerous frequency channels and improved modelling will help characterising and removing other CMB foregrounds \citep{Maniyar2021, Douspis2022, Raghunathan2023}. 

\section{Data}
\label{sec:data}

\begin{figure}
    \centering
    \includegraphics[width=\columnwidth]{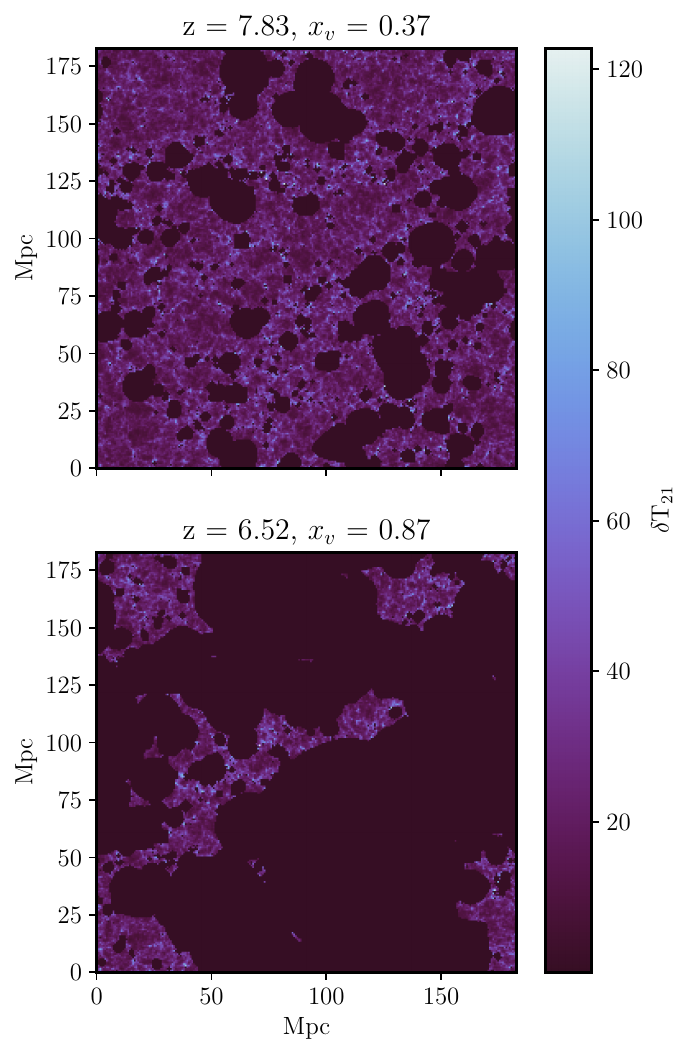}
    \caption{ Slices of the differential surface brightness temperature $\delta T_{21}$ for the {\sc RSAGE} simulation. The upper panel is at a redshift of $z = 7.83$ and has a volume-averaged ionisation fraction of $x_v =  0.37$, while for the bottom panel, the same values are $z = 6.52$ and $x_v = 0.87$. Both redshifts correspond to the mock data points used in Section~\ref{subsec:3_fit}.
    }
    \label{fig:rsage_dt_slices}
\end{figure}

We briefly describe the {\sc RSAGE} simulation used in this work \citep[see ][for a detailed description]{Seiler2019} to validate our simplified parameterisation of the 21\,cm power spectrum given in equation~\eqref{eq:recons_21cm_ps_simplified}. The $N$-body data underlying {\sc RSAGE} has been generated with the hydrodynamic {\sc Kali} code \citep{Seiler2018} with $2400^3$ dark matter particles and a volume of (160\,Mpc)$^{3}$. The radiative transfer is conducted using the semi-numerical code {\sc CIFOG} \citep{Hutter2018}, accounting for star formation and feedback processes. We use the \texttt{const} version of the simulation, which assumes a constant escape fraction of ionising photons $f_{\mathrm{esc}} = 0.2$. In Fig.~\ref{fig:rsage_dt_slices}, we show simulation slices of the differential surface brightness temperature of the 21\,cm signal. The cosmology used is consistent with \citet{PlanckCollaborationAghanim_2020} and summarised in Table~\ref{tab:rsage_ref_params}. 
The reionisation mid- and endpoint values given in Table~\ref{tab:rsage_ref_params} are obtained by fitting the asymmetric parameterisation from equation~\eqref{eq::reion_parametrisation} to the reionisation history of the simulation. Similarly, the values of the morphology parameters log$_{10}(\alpha_{0})$  and $\kappa$ from Table~\ref{tab:rsage_ref_params} are derived by fitting the {\sc RSAGE} electron density power spectrum with equation~\eqref{eq:Pee_model} \citep[see][for details of the fit]{GorceIlic_2020}. We have checked that the baryon power spectrum from {\sc RSAGE} is well described using equation~\eqref{eq::baryon_ps} for the given values of $k_f$ and $g$. Overall, we find the model of the electron density power spectrum in equation~\eqref{eq:Pee_model} to be a good fit for the redshift and $k$-scales explored in this work. In the following section, we therefore limit our inspections to the accuracy of the 21\,cm power spectrum reconstruction.

\section{Results}
\label{sec:results}

In this section, we first look at the model uncertainties associated with our simplified derivation of the 21\,cm power spectrum from equation~\eqref{eq:recons_21cm_ps_fit} given an electron power spectrum by comparing it to the full spectrum computed from the {\sc RSAGE} simulation. In Sec.~\ref{subsec:3_fit}, we derive forecast constraints on reionisation by applying this derivation to fit mock 21\,cm and kSZ data points, using equations~\eqref{eq:recons_21cm_ps_fit} and~\eqref{eq:Pee_model}, to a reionization model with the parameters from Table~\ref{tab:rsage_ref_params}. We showcase the advantages of combining the two datasets compared to a separate analysis. In appendix~\ref{appendix::bias}, we extend this analysis with 21\,cm and kSZ data for the from {\sc RSAGE} simulation, to study the effect of the model assumptions.

\subsection{Reconstructing the power spectra}
\label{subsec:3_recons}

\begin{figure}
    \centering
    \includegraphics[width=\columnwidth]{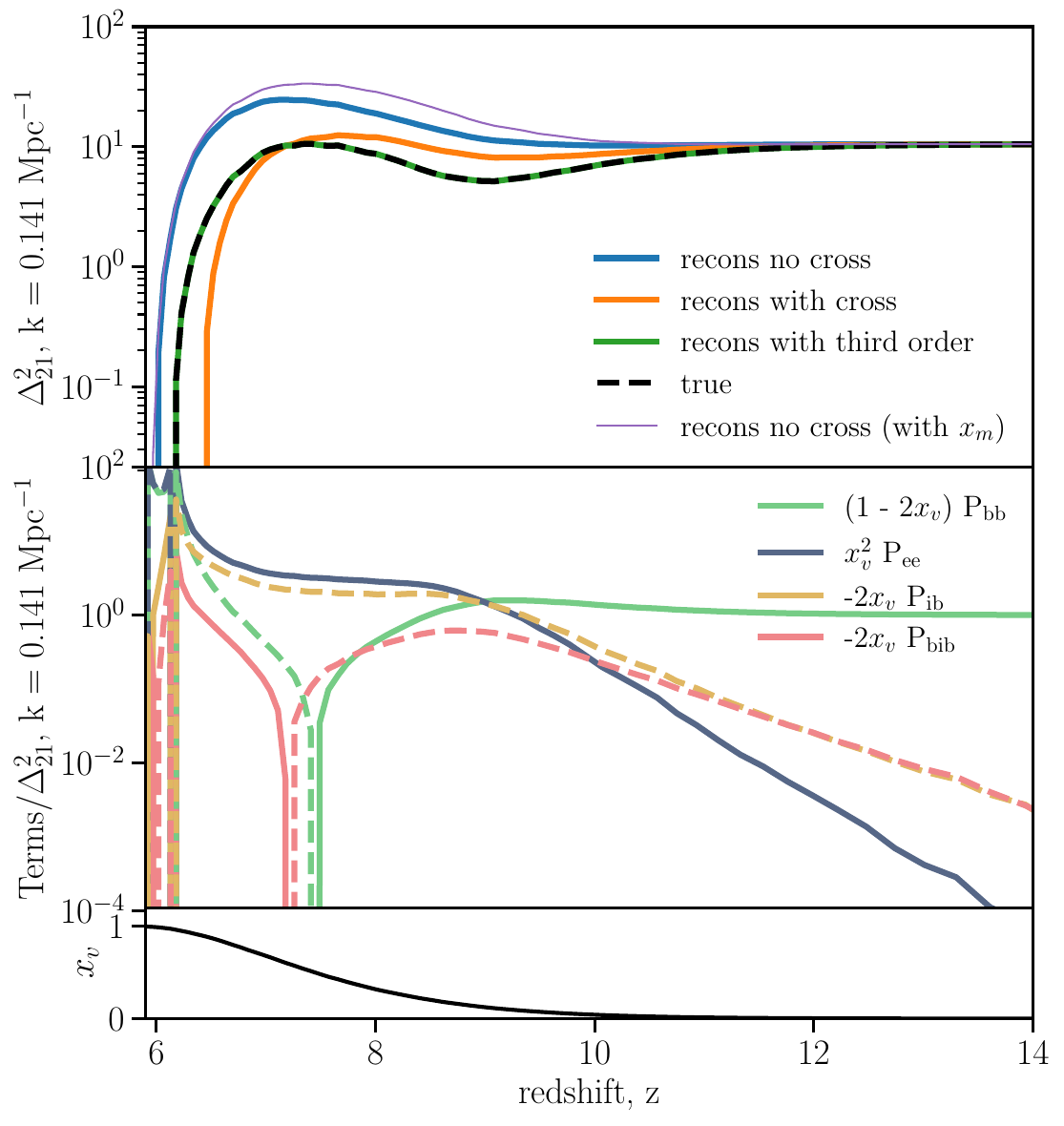}
    \caption{~\textit{Upper Panel}: The dimensionless 21\,cm power spectrum at $k = 0.141~\mathrm{Mpc}^{-1}$ as a function of redshift. The power spectrum computed directly from the {\sc RSAGE} simulation is represented by the black dashed line. The coloured lines correspond to different reconstruction levels (see text for details). ~\texttt{recons no cross} corresponds to equation~\eqref{eq:recons_21cm_ps_fit}, which is used for the forecast in section~\ref{subsec:3_fit}. Additionally,~\texttt{recons no cross (with $x_m$)} corresponds to equation~\eqref{eq:recons_21cm_ps_fit} but for $x_m \neq x_v$. In ~\texttt{recons with cross}, the cross power spectrum is included: $-2x_v(z) P_{bi}(k,z)$. Lastly, ~\texttt{recons with third order} adds three-point power spectrum contribution to ~\texttt{recons with cross}: $ -2x_v(z) P_{bi,b}(k,z)$, which corresponds to the full form of the 21\,cm power spectrum in equation~\eqref{eq:recons_21cm_ps_full}.
    ~\textit{Middle Panel}: Ratio of each component in equation~\eqref{eq:recons_21cm_ps_full} against the total 21\,cm power spectrum. In this panel, solid (dashed) lines represent a positive (negative) contribution. 
    \textit{Bottom Panel}: Reionization history of the {\sc RSAGE} simulation.
    }
    \label{fig:recons_comp}
\end{figure}

We want to understand the role of each contributing term in equation~\eqref{eq:recons_21cm_ps_full}, especially the ones we have discarded in our simplified $P_{21}$ reconstruction (equation~\ref{eq:recons_21cm_ps_fit}), in order to assess its accuracy. We reconstruct and examine the 21\,cm power spectrum from the {\sc RSAGE} simulation described in section~\ref{sec:data}. The top panel of Fig.~\ref{fig:recons_comp} showcases the true $P_{21}$ plotted against different reconstruction precision levels:
\begin{description}
    \item[-] The \texttt{recons no cross} case corresponds to the highest simplification level, corresponding to equation~\eqref{eq:recons_21cm_ps_fit}, and the approximation we will use in the remaining of this work. Here, it is computed with the true $P_{ee}(k,z)$ spectrum obtained from the sim ulation, and not equation~\eqref{eq:Pee_model}, therefore, we only assess the accuracy of the 21\,cm reconstruction. The contributions from both third-order terms and cross-terms are discarded. We also include an additional case, \texttt{recons no cross (using $x_m$)}, where we lift the assumption of equating the mass-weighted and volume-weighted ionization fractions.
    \item[-] For \texttt{recons with cross}, we add the cross power spectra contributions: $-2x_v(z) P_{bi}(k,z)$\footnote{\label{note1}Note that this contribution can be positive or negative, as illustrated in Fig.~\ref{fig:recons_comp}.},
    \item[-] For \texttt{recons with third order}, we further add the three-point power spectrum contribution: $ -2x_v(z) P_{bi,b}(k,z)$\footref{note1}. This corresponds to the full form of the 21\,cm power spectrum, without simplification.
\end{description}

We look at the relative contribution of each of the terms in equation~\eqref{eq:recons_21cm_ps_full} to the 21\,cm power spectrum in the {\sc RSAGE} simulation. This is represented in the middle panel of Fig.~\ref{fig:recons_comp} for $k = 0.141~\mathrm{Mpc}^{-1}$. %The solid lines represent a positive contribution, while the dotted lines are the absolute values of negative contributions. 
We see that for $z>10$, in the very early stages of EoR, most of the 21\,cm power spectrum amplitude comes from the matter density. In this period, despite the low ionisation level ($x_v \lesssim 10^{-4}$), the ionisation and density fields are weakly correlated, and the role of higher-order terms is negligible \citep{LidzZahn_2007}: All reconstruction levels of $P_{21}$ look identical in the upper panel. Beyond $ x_v \approx 10 \%$, when the first luminous sources have reionised their local over-densities and the ionisation field is correlated with the density field, $P_{21}$ is primarily determined by the difference between the electron density power spectrum and the $P_{bi}$ cross-spectrum. %\ag{\st{The electron density power spectrum $P_{ee}$ is always positive compared to the cross-power spectrum $P_{ib}$, which is negative on large scales.}} 
The third-order term $P_{bib}$, in red, plays a minor role during most of reionization, only gaining significance in the final stages. This term is initially negative and changes sign at the midpoint of reionization because of the change in correlation between the density and ionisation fields \citep{GeorgievMellema_2021}. Conversely, for the density term, the change in sign is due to the $[1-2x_v(z)]$ factor present in equation~\eqref{eq:recons_21cm_ps_simplified}. %Note that both terms are similar in magnitude and opposite in sign within the second half of reionization.

We can form a clearer understanding of the impact of each term by examining the top panel of Fig.~\ref{fig:recons_comp}, where the power spectrum of the 21\,cm signal is calculated according to equation~\eqref{eq:recons_21cm_ps_fit}, in blue, while the full model presented in equation~\eqref{eq:recons_21cm_ps_full}, in green. %The orange line is then the same equation with the addition of the cross power, while the green line also considers the cross power and higher order term. The thin black line showcases the reionization history of the globally averaged ionisation fraction against redshift.
%In the simple case where we do not consider any of the additional power spectra, there is an evolving positive bias throughout the EoR. 
We see that the simplified expression in the ~\texttt{recons no cross} model overestimates the true power throughout reionization. Early on, ~\texttt{recons no cross} follows the true~\texttt{recons with third order} model. With the onset of reionization, a positive bias of an order of two appears between the true spectrum and its model, reaching a global maximum at the midpoint of reionization. %Additionally, the simplified model is incapable of reproducing the feature around $z\sim9$ when the cross and density power spectra are of roughly equal amplitude (we refer the interested reader to section~2.2 of \citet{LidzZahn_2008} for a more detailed discussion of such effect).
Additionally, in the simplified model, reionization is delayed by four per cent compared to the true model. Hence, our simplified model will tend to overestimate $P_{21}$ and will be biased towards late reionization scenarios, the implications of which are discussed in section~\ref{subsec:3_fit}. Additionally, we note that by removing the $x_v \equiv x_m$ assumption and replacing $x_v^2 P_{ee}$ by $x_m^2 P_{ee}$ in equation~\eqref{eq:recons_21cm_ps_simplified} results in a higher amplitude (seen in purple) compared to the \texttt{recons no cross} model (light blue) using the volume-weighted fraction. The increased bias of the model is due to the mass-weighted ionisation fraction being larger than the volume-weighted ionisation fraction $x_v > x_m$ at all redshifts in inside-out models of reionisation \citep[see, e.g.,][]{Dixon2016}. We have compared this across the $k$-scales accessible within the {\sc RSAGE} simulation volume, and we find the $x_v \equiv x_m$ simplification leads to a $P_{21}$ model closer to the truth when excluding the higher-order contributions.

Adding only the contribution of the cross-power spectrum in the calculation (\texttt{recons with cross}) does not help the reconstruction as much as one would hope: While the additional negative power from the cross-correlation decreases the bias of the estimate to that of the true power spectrum, the exclusion of the higher-order power spectrum shifts the midpoint redshift and biases the result to an early end of reionization.

The role of the higher-order terms in the derivation of the 21\,cm power spectrum is non-trivial and varies across the $k$-scales and redshifts. % (see Fig.~\ref{fig:recons_comp_extra} for a figure similar to Fig.~\ref{fig:recons_comp} but at $k=0.5\,\mathrm{Mpc}^{-1}$).
%\citep[e.g., figures~2 \&  4 of][]{LidzZahn_2007}.
Generally, the negative cross-correlation between the density and ionisation fields weakens at higher $k$-scales as well as with the progression of reionization. %The role of the cross-power spectrum diminishes, while the significance of the higher-order power spectrum rises. 
We see that by excluding the higher-order power spectrum and varying the $k$-scale, the \texttt{recons with cross} model converges to the \texttt{recons with no cross} model on large $k$-scales. In Appendix~\ref{app:obs_reconstruction}, we look at the possibility of using equation~\eqref{eq:recons_21cm_ps_simplified} to reconstruct the pkSZ angular power spectrum from an interferometric measurement of the 21\,cm power spectrum.

In summary, the different terms comprising our expression of the 21\,cm power spectrum, described in equation~\eqref{eq:recons_21cm_ps_full}, are non-trivial. The importance of each term presented in Fig.~\ref{fig:recons_comp} is shown to evolve with redshift and $k$-scale. The model corresponding to the highest degree of simplification is the most promising for this work, as the amplitude of the simplified 21\,cm power spectrum is positively biased on all $k$-scales, resulting in a positive bias on the end of reionization, which we can easily quantify. In contrast, including the cross-power spectrum $P_{bi}$ contribution to the model introduces an evolving amplitude bias with $k$-scale. Hence, we choose to perform our forecast with the \texttt{recons no cross} model, whilst keeping its limitations in mind. We use this model to derive our true mock data points and each sampled model. 

For the sake of clarity, we list below the different layers of assumptions for our model in equation~\eqref{eq:recons_21cm_ps_fit} and how we address them.

\begin{enumerate}
    \item The baryon power spectrum follows equation~\ref{eq::baryon_ps} when using the {\sc CAMB}-derived theory non-linear matter power spectrum. Comparing to the spectrum directly measured from {\sc RSAGE}, we find a good match with the model across all redshifts and $k$-scales covered in this work.
    \item The global reionisation history follows equation~\ref{eq::reion_parametrisation}. We have compared the evolution obtained with equation~\ref{eq::reion_parametrisation} and the parameters of Table~\ref{tab:rsage_ref_params} to the {\sc RSAGE} history seen at the bottom panel Fig.~\ref{fig:recons_comp} and found to be a good match for the redshifts considered in this paper.
    \item The mass- and volume-weighted ionisation fractions $x_m$ and $x_v$ are identical. As seen in the upper panel of Fig.~\ref{fig:recons_comp}, the positive bias induced by excluding the higher order terms is slightly alleviated for the \texttt{recons no cross} model for the inside-out reionization models considered in this work, for which $x_v < x_m$.
    
    \item The true electron power spectrum in {\sc RSAGE} follows equation~\eqref{eq:Pee_model}. This model has been compared against the {\sc RSAGE} simulation within this work while the best-fit morphology parameters are reported in Table~\ref{tab:rsage_ref_params} \citep[see appendix~B.2 of][for a detailed discussion]{GorceIlic_2020}.
\end{enumerate}

\begin{figure}
    \centering
    \includegraphics[width=\columnwidth]{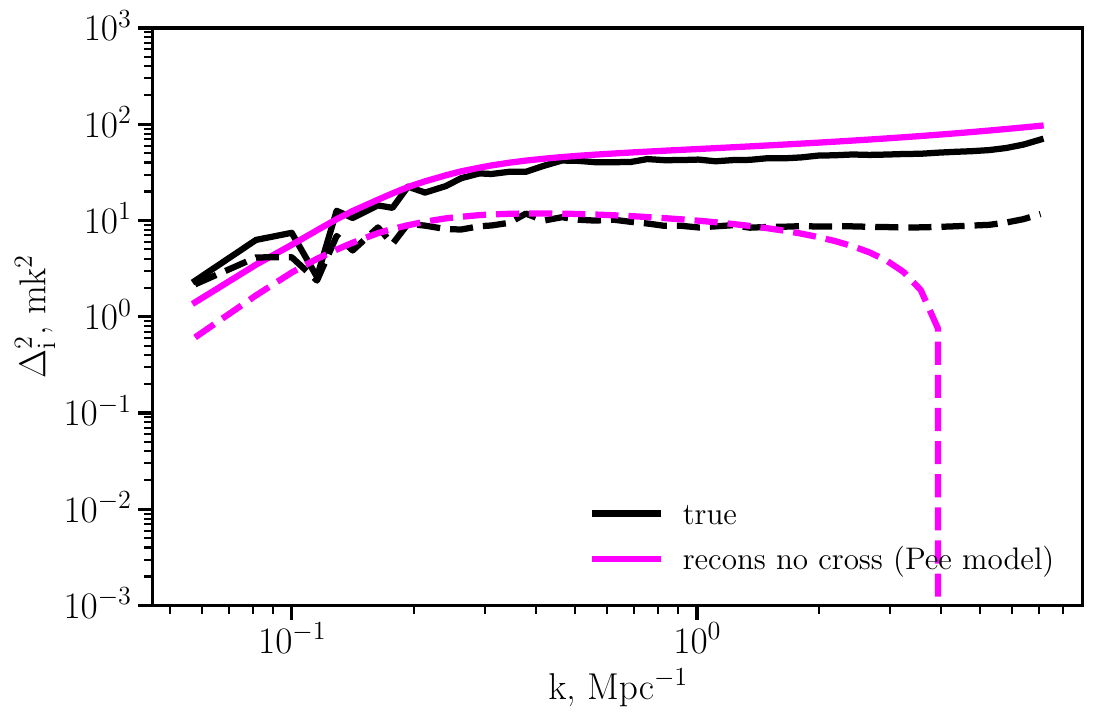}
    \caption{The 21\,cm power spectrum {\sc RSAGE} (seen in black) compared against our model (seen in fuchsia) for redshifts $z = 6.5, 7.8$ ($x_{v} = 0.87, 0.37$) as thick solid and dashed lines, respectively. Our model 21\,cm power spectrum has been generated using equation~\eqref{eq:recons_21cm_ps_fit} for the reference parameters in Table~\ref{tab:rsage_ref_params} and the {\sc CAMB} non-linear matter power spectrum.}
    \label{fig:model_compare}
\end{figure}

We conclude this assessment of our model uncertainties by comparing, in Fig.~\ref{fig:model_compare}, the true 21\, power spectrum from {\sc RSAGE} against the one generated by our fully simplified parameterisation, given in equation~\eqref{eq:recons_21cm_ps_fit}. Overall, the parameterisation performs remarkably well on the $k$-scales considered in this work ($k\sim0.1~\mathrm{Mpc}^{-1}$). At high $k$-scales ($k > 1.0$ \,Mpc$^{-1}$), the model in equation~\eqref{eq:Pee_model} scales as $k^{-3}$ and
quickly approaches the baryon power spectrum. The result is a negative 21\,cm power spectrum as can be observed in Fig.~\ref{fig:model_compare} at z = 6.5. Therefore, the model should not be used as is at very small scales. At low-$k$, differences between the model and the {\sc RSAGE} power spectrum will be sensitive to sample variance due to the simulation box size and hence have not been studied within this work.

We discuss the effect of the aforementioned bias on our results in section~\ref{appendix::bias}, where we derive the mock data using the {\sc RSAGE} simulation and no approximation. In Sec.~\ref{sec:conclusions}, we discuss future improvements and possible avenues in modelling the higher-order terms in our model.

\subsection{Efficient probe combination}
\label{subsec:3_fit}
In this section, we fit the midpoint, endpoint and morphology parameters of reionization to three mock data points: One measurement of the pkSZ angular power spectrum at multipole $\ell=3000$ and two measurements of the 21\,cm power spectrum. The choice of the data points is motivated by current upper limits on both observables \citep[e.g.,][]{Trott2020, ReichardtPatil_2021, GorceDouspis_2022}, as well as our estimate of measurement errors from section~\ref{subsec:2_errors}. The number of data points used is chosen to efficiently retrieve information on the EoR, whilst intuitively illustrating the role of each measurement. Limiting the number of data points will also help us understand how many observations and of what quality are necessary to begin constraining the properties of reionization, in a context of gradual improvement of upper limits on the 21\,cm power spectrum. In section~\ref{sec:discussion}, we investigate how the quality of constraints depends on the redshift and scale chosen for our data points. We first forecast results obtained with 1000\,hrs of SKA-\textit{Low} observations before turning to mock observations by MWA and LOFAR.

\subsubsection{Forecast with SKA-Low}
\label{sec::SKA_ideal}

\begin{figure}
    \includegraphics[width=1.\columnwidth]{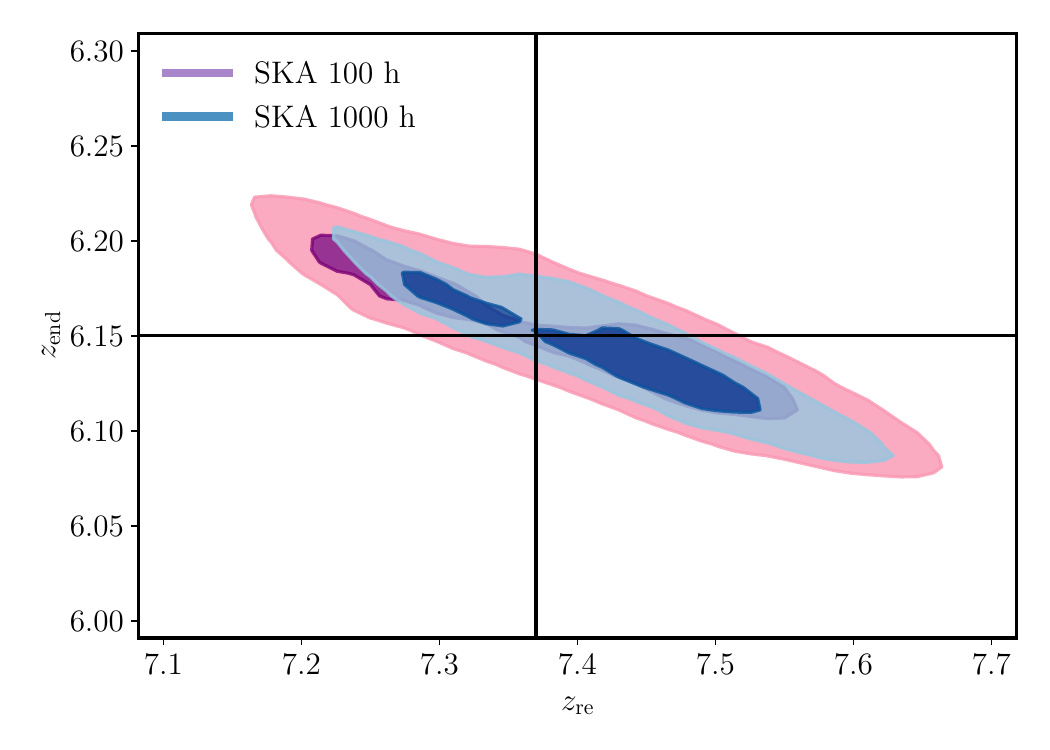}
    \includegraphics[width=1.\columnwidth]{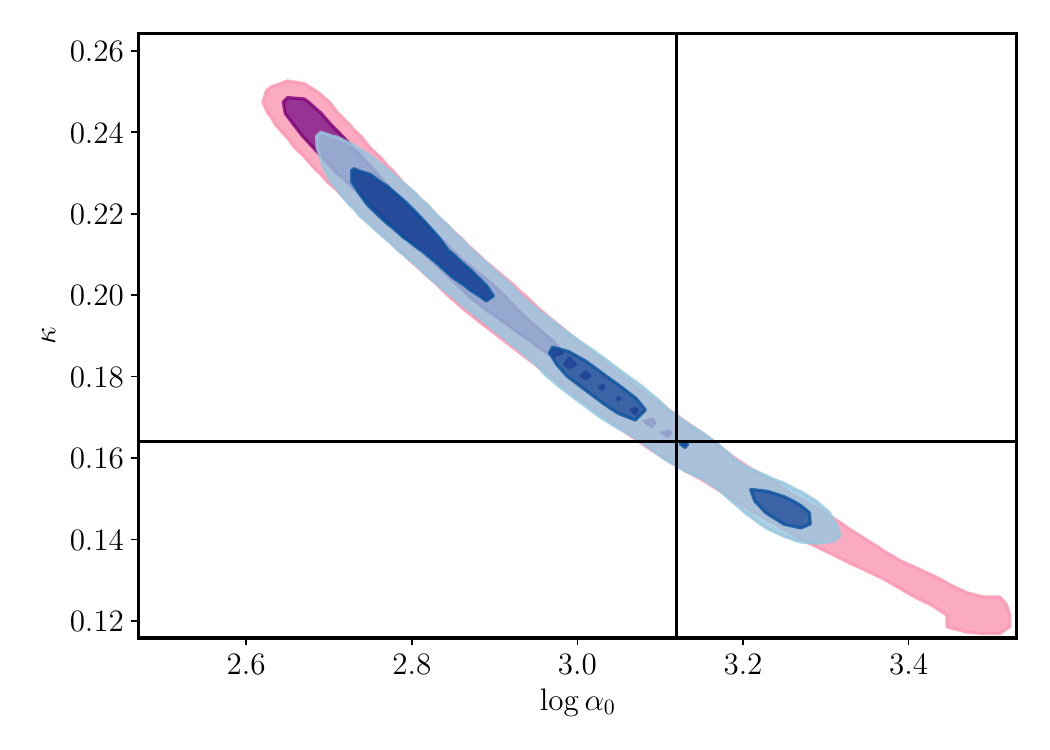}
    \caption{Posterior distributions on reionization mid- and endpoint (upper panel) and the morphology parameters log$_{10}(\alpha_{0})$ and $\kappa$ (bottom panel), considering a detection of the pKSZ power spectrum at $\ell = 3000$ and of the 21\,cm power spectrum at redshifts $z =6.5,7.8$ and $k = 0.50$\,Mpc$^{-1}$ for either 100 and 1000~hours of integration with SKA-\textit{Low}, in purple and in blue, respectively. Note that the axis ranges in both panels are smaller than the priors in Table~\ref{tab:rsage_ref_params} or the ones presented in following figures.}
    
    \label{fig::SKA_ideal}
\end{figure}

Our primary scenario is based on detection of the 21\,cm power spectrum at redshifts $z =6.5$ and $7.8$, both at $k = 0.50$\,Mpc$^{-1}$, that is outside the foreground wedge \citep{LiuParsons2014}, considering noise levels associated with 100 and 1000~hours of observations with the SKA-Low. Additionally, a measurement of the pkSZ power spectrum is assumed at $\ell = 3000$ with a ten per cent error bar such that $D^{pkSZ}_{l =3000} = 0.86 \pm 0.09 \, \mu K^{2}$. For simplicity, we will refer to these cases as \texttt{1k2z}, based on the choice of two observations of the 21\,cm power spectrum at the same $k$-scale.

In Fig.~\ref{fig::SKA_ideal}, we show the joint probability distributions for our four sampled parameters (see appendix~\ref{appendix::3k2z} for a discussion on parameter degeneracies). The best-fit parameters, their 1$\sigma$ error, and Gelman-Rubin convergence parameters are presented in Table~\ref{tab:MWA_MCMC_results}\footnote{The Gelman-Rubin parameters for morphology parameters of the 1000 h case, given in Table~\ref{tab:MWA_MCMC_results}, are of order $R-1 \approx 1$, so that the chains are not fully converged despite running for over a million iterations. However, the worst values are obtained for parameters we marginalise over, and we noticed that the posteriors of relevant parameters do not change significantly as iterations increase, confirming that the results still hold good qualitative significance.}.

Our results indicate that we can recover the true values of $z_{\mathrm{re}}$ and $z_{\mathrm{end}}$ in both the \texttt{1k2z} cases, despite the order of magnitude difference in integration time between them. Moreover, we independently constrain the value of the Thomson optical depth with a deviation of less than one per cent of the true value, $\tau = 0.066 \pm 0.002$ (see Sec.~\ref{appendix::tau} for a more detailed analysis). Therefore, even early on in the operation of SKA-\textit{Low}, we will gain insight into the cosmological properties of reionization with minimal knowledge of the properties of ionising sources (marginalising over them). By fixing the $k$-scale and varying the redshift, the forecast is naturally sensitive to the variance of the 21\,cm field. This is partly due to the choice of redshift, as once significant overlap of ~\HII bubbles has taken place, that is past the midpoint of the EoR, the 21\,cm power spectrum decreases in amplitude. %To test this we have also performed an additional \texttt{1k2z} test where we fix $k = 0.1$\,Mpc$^{-1}$ and see a decrease in the constraining power of the forecast for $z_{\mathrm{re}}$ \& $z_{\mathrm{end}}$. 

%Because the break feature indicative of the characteristic scale of ~\HII bubbles moves to lower-$k$ of the 21\,cm power spectrum with the EoR \citep{Furlanetto2006}, the choice of a fixed $k$-scale will result in less constraining power on the astrophysical parameters  log$_{10}(\alpha_{0})$ \& $\kappa$.
For both cases presented in Fig.~\ref{fig::SKA_ideal}, we can place loose lower limits on  log$_{10}(\alpha_{0})$, although the 1000\,hrs case is more constraining on both distributions: Adding integration time additionally provides a lower limit on $\kappa$. Note that the ranges of the axes in the figure are smaller than those of the priors from Table~\ref{tab:rsage_ref_params} (while the limit on $\kappa$ is extended), confirming that these limits are not only informed by the priors. We further investigate whether the choice of $k$-scale has an effect on these results in section~\ref{subsec::2k1z}.\\

% Below lies the old text for the 2 parameter fit, revive if necessary
%However, note the skew in the individual posterior distributions of $z_{\mathrm{re}}$ and $z_{\mathrm{end}}$, which results in a positive correlation between both values in the two-dimensional posterior. We explore two possible methods to reduce the bias. In the first case we increase the integration time by an order of magnitude and thus reducing the error of the 21\,cm measurement as $\Delta_\mathrm{noise}^{2} \propto 1/t_\mathrm{int}$. This results in a more accurate recovery of the true values, albeit the bias is not completely removed. On the other hand, we choose and integration time to 100\,hours and a prior on the Thomson optical depth. The choice of prior indirectly excludes the late and rapid models introduced from the bias due to their low $\tau$ value. While the true values is within the $1\sigma$ uncertainty in both of these cases, we find that the most effective method of eliminating the bias is by decreasing the noise estimate of the 21\,cm power spectrum measurement. \ig{Might be worth mentioning, I have run a SKA 100h integration time and can do for HERA. The noise is so  much lower that they will be able to recover the true value with no bias.}  

\begin{figure*}
    \centering
    \includegraphics[width=\textwidth]{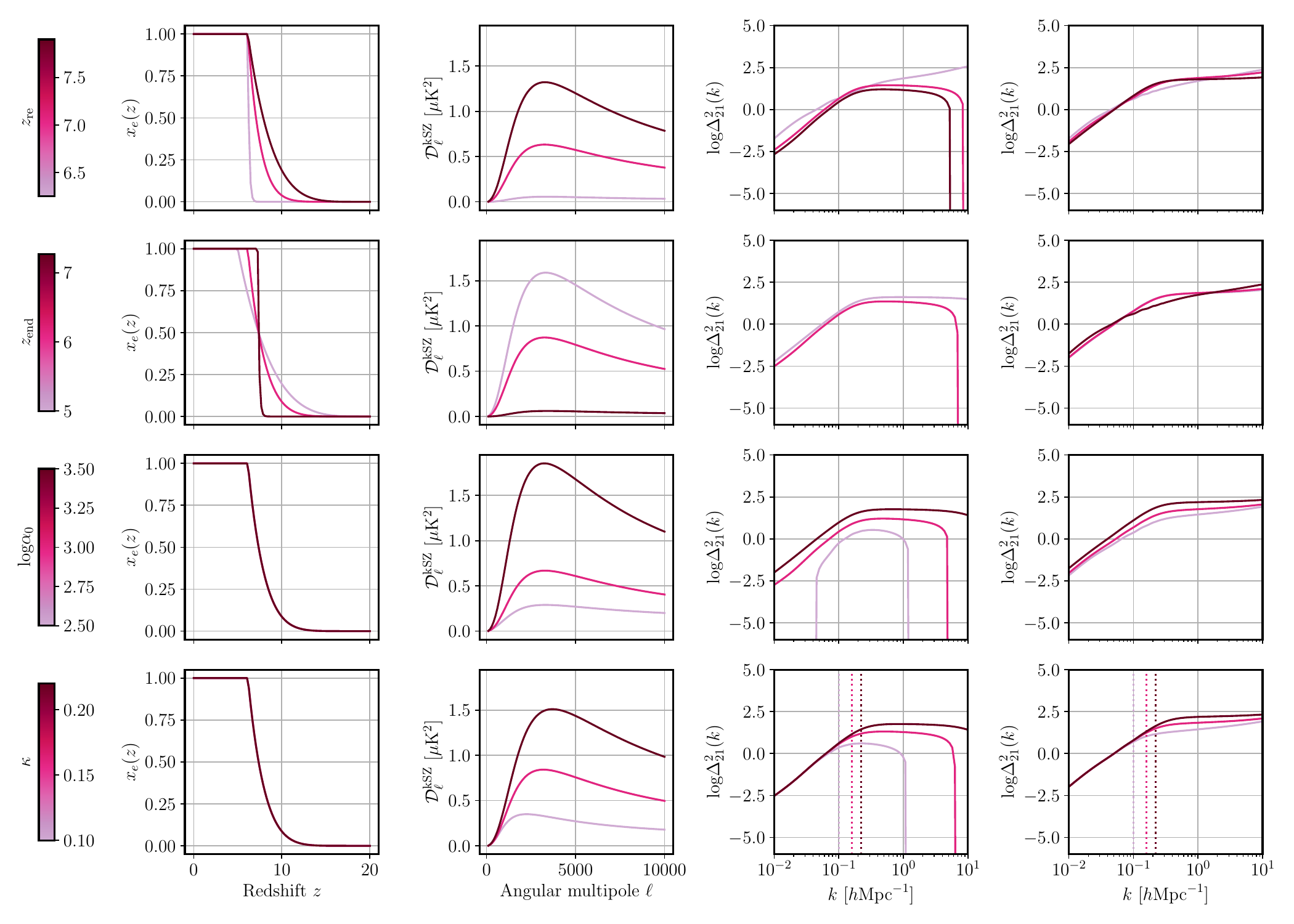}
    \caption{Evolution of the reionization history (first column), the pkSZ angular power spectrum (second column), and the dimensionless 21\,cm power spectrum at redshifts $z = 6.5, 7.8$, considered in section~\ref{subsec:3_fit} (last two columns, respectively) as a function of our four model parameters. The values of $\kappa$ are shown as vertical dotted lines in the final row with the corresponding colour. Note the difference in the shape of the 21\,cm power spectrum at high-$k$ compared to Fig.~\ref{fig:tel_noise} is discussed in Sec.~\ref{subsec:3_recons}.}
    \label{fig:app_params_vs_data}
\end{figure*}

\begin{table*}
    \centering
    \begin{tabular}{lcccccccc}
        Models & & &  $z_{\mathrm{re}}$ & $z_{\mathrm{end}}$ & log$_{10}(\alpha_{0})$ & $\kappa$ & $\tau$ & d$z$  \\
        \hline\hline
        Label & Data & True & 7.37  & 6.15 &  3.12 & 0.16 & 0.0649 & 1.22  \\
        \hline
        \texttt{1k2z} & z = 6.5, 7.8 &1000h  & 7.39 $\pm$ 0.14  & 6.15 $\pm$ 0.04  & 3.04 $\pm$ 0.32 &  0.18 $\pm$ 0.04  & 0.0651 $\pm$ 0.0018  & 1.24 $\pm$ 0.17\\
        &k = 0.5\,Mpc$^{-1}$ &R - 1  & 0.3  & 1.5  & 1.6 &  0.5  & N/A  & N/A\ \\
        \hline 
        \texttt{1k2z}&  z = 6.5, 7.8  &100h  & 7.42 $\pm$ 0.13  & 6.15 $\pm$ 0.04  & 2.95 $\pm$ 0.31 &  0.19 $\pm$ 0.04  & 0.0655 $\pm$ 0.0017  & 1.27 $\pm$ 0.16\\
        & k = 0.5\,Mpc$^{-1}$ &R - 1  & 0.1  & 0.1  & 0.3 &  0.3  & N/A  & N/A\ \\
        \hline 
    \end{tabular}
    \caption{Best fit values of the model parameters for the \texttt{1k2z} cases, their corresponding 1$\sigma$ uncertainty and Gelman–Rubin parameter. Best-fit values and errors are also given for derived parameters such as $\tau$.}
    \label{tab:MWA_MCMC_results}
\end{table*}
To get more insight into the results above, we look in more detail at the dependence of our two observables on the model parameters. Fig.~\ref{fig:app_params_vs_data} presents the reionization history, the pkSZ power spectrum, and the 21\,cm power spectrum at $z=6.5$ and 7.8, corresponding to the two redshifts considered above for the mock data points (in each column of the figure, respectively), for different values of the model parameters. All parameters are varied within the value ranges used as flat priors in the analysis.  
Regarding the global reionization parameters, we see that varying the mid- and endpoint leads to similar, although opposite, variations in both observables: Increasing $z_\mathrm{re}$ extends the duration of the EoR, which leads to a boosted amplitude of the pkSZ power spectrum while having a smaller effect on the amplitude of the 21\,cm power spectrum. Conversely, increasing $z_\mathrm{end}$ decreases the duration of the EoR, suppressing the pkSZ amplitude. This explains the strong anti-correlation observed between the two parameters in Fig.~\ref{fig::SKA_ideal}. Note that for the model where $z_\mathrm{end}=7$, naturally, the 21\,cm power spectrum is zero at $z=6.5$.

On the other hand, $\log_{10}(\alpha_0)$ and $\kappa$ are also strongly correlated. The morphology parameters have no influence on the global reionization history but impact the shape of the electron power spectrum and, in turn, the shape and amplitude of the 21\,cm and pkSZ power spectra. However, compared to the global history parameters, increasing $\log_{10}(\alpha_0)$ or $\kappa$, as seen in the third and fourth columns of Fig.~\ref{fig:app_params_vs_data}, results in a similar impact on both observables. The $\log_{10}(\alpha_0)$ parameter is effectively a measurement of the large-scale amplitude of $P_{ee}$ at high redshift (before the reionization midpoint), whilst the $P_{ee}$ power breaks as $k^{-3}$ for scales $k >\kappa$, such that $\kappa$ can be interpreted as the minimal size of ionised regions during reionization. Hence, varying log$_{10}(\alpha_{0})$ directly impacts the amplitude of both observables, whilst varying $\kappa$ changes the shapes of both power spectra: The maximum of the pkSZ power and the shoulder in the dimensionless 21\,cm power spectrum shift towards smaller scales as $\kappa$ increases. This dependence on the observables on these two parameters explains the degeneracy between the morphology parameters seen in Fig.~\ref{fig::SKA_ideal}. For the \texttt{1k2z} case, where both data points are at a fixed $k$-scale, a lower value log$_{10}(\alpha_{0})$ can be compensated by a higher $\kappa$ value, within the measurement error of the data. However, this degeneracy can be reduced in the presence of 21\,cm data points at different $k$-scales or by kSZ data points at different multipoles, since $\kappa$ impacts the shape of both power spectra. We explore the $k$-scale dependency of our chosen data in section~\ref{subsec::2k1z}.

\subsubsection{Constraining power of each dataset}

\label{secc:each_data_set}
\begin{figure}
    \centering
    \includegraphics[width=1.\columnwidth]{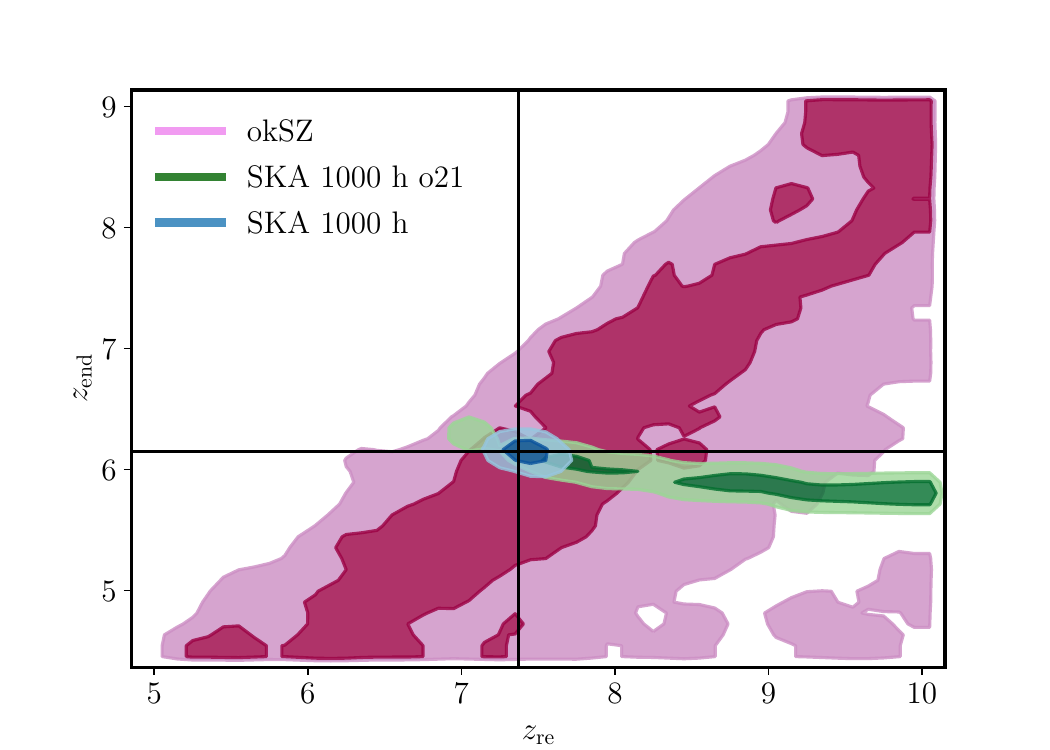}
    \includegraphics[width=1.\columnwidth]{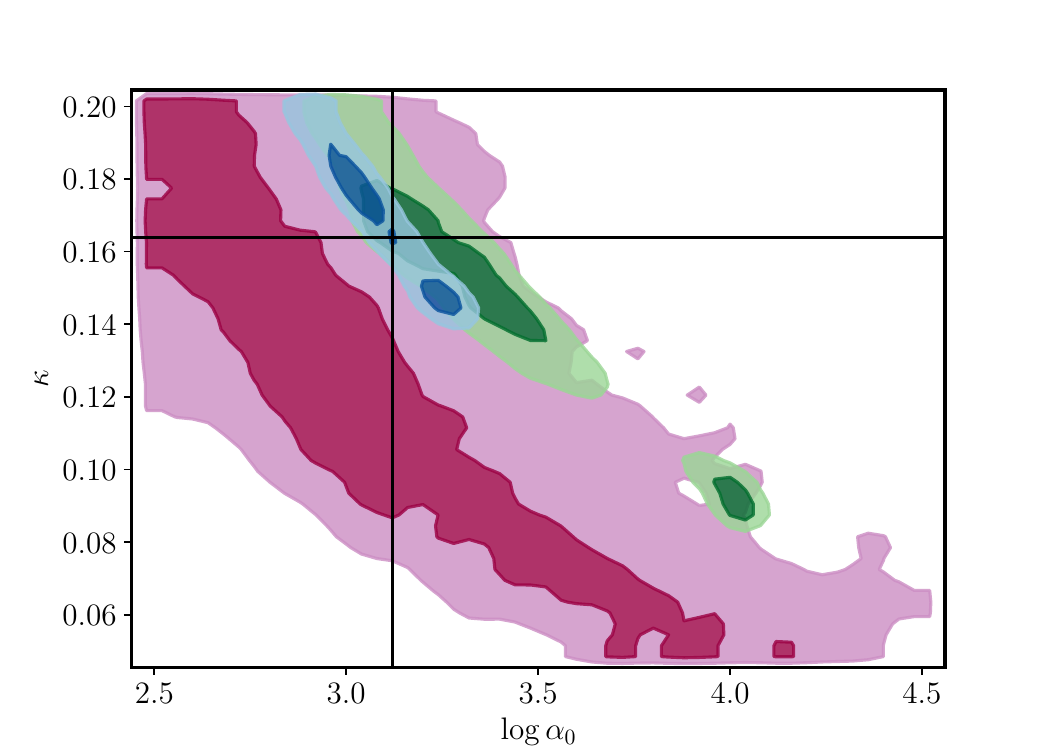}
    \caption{ Posterior distributions on reionization mid- and endpoint (upper panel) and the morphology parameters log$_{10}(\alpha_{0})$ and $\kappa$ (bottom panel) when fitting 21\,cm and pkSZ power spectra individually (green and magenta) or jointly (in blue), for 1000~hours of SKA-\textit{Low} observations. The vertical and horizontal lines show the `true' values used to generate the mock data.}
    \label{fig:zre_zend_comparison_data}
\end{figure}

In this subsection, we examine each dataset's ability to individually impose constraints on reionization.
In Fig.~\ref{fig:zre_zend_comparison_data}, we present the joint posterior distribution of the sampled parameters for three different cases assuming a SKA-\textit{Low} detection after integrating over 1000\,hours. We compare a case where the forecast is run only with the 21\,cm data points (\texttt{o21}), a case with the pkSZ measurement only (\texttt{opkSZ}), and, lastly, the joint forecast previously illustrated in Fig.~\ref{fig::SKA_ideal}.
To better understand the role of each parameter in each of the aforementioned cases, we firstly group our analysis by examining the global reionization parameters $z_\mathrm{re}$ and $z_\mathrm{end}$, then the astrophysical log$_{10}(\alpha_{0})$ and $\kappa$, in Fig~\ref{fig:zre_zend_comparison_data}.

Examining the two-dimensional $z_\mathrm{re}$ and $z_\mathrm{end}$ posterior distributions for the \texttt{opkSZ} case, we find that the parameters are positively correlated. As discussed in section~\ref{subsec:3_fit}, this occurs because the pkSZ effect is sensitive to the duration of reionization $\mathrm{d}z\equiv z_\mathrm{re}-z_\mathrm{end}$ as the rate of the growth of ionising bubbles impacts the amplitude of the signal (see also the first and second column of Fig.~\ref{fig:app_params_vs_data}). Models with an extended reionization period result in a larger amplitude of the pkSZ power spectrum as more free electrons interact with the ionised medium, and reciprocally \citep{McQuinn2005,Mesinger2012,BattagliaTrac_2013}. Note that the posterior distribution for the \texttt{opkSZ} case covers a region of parameter space where the universe has reionised earlier than the redshift at which our 21\,cm data points are measured. These models are naturally excluded when combining both data sets. For the \texttt{o21} case, the data constraints tightly constrain the end of reionization but only place an upper limit on the midpoint. Compared to the pkSZ-only case, the \texttt{o21} case is not sensitive to the duration of reionization and favours extended models of the EoR, as previously noted by \citet{BeginLiu_2022}. Conversely, a detection of the 21\,cm power spectrum at $z = 6.5$ naturally excludes models where reionization finishes earlier than that redshift, hence, the forecast gives a lower limit on $z_{\mathrm{end}}$. Lastly, models at the lower left of the two-dimensional posterior ($z_{\mathrm{re}} \lesssim 7$ and $z_{\mathrm{end}} \lesssim 6$) are excluded by the 21\,cm data at $z = 6.5$ and 7.7. For these models, the 21\,cm power spectrum at these redshifts would trace the density power spectrum. Such a feature is inconsistent with the amplitude of the 21\,cm data points and their associated error bars. 
Therefore, the complementary nature of both probes breaks the degeneracy inherent to each data set and allows us to constrain the parameter space and recover the true reionization history. 

Regarding log$_{10}(\alpha_{0})$ and $\kappa$, we see that both morphology parameters are anti-correlated and the pkSZ data alone only provides a biased measurement. This can be better understood by examining the two lower rows of Fig.~\ref{fig:app_params_vs_data}: Increasing either of the morphology parameters results in a boosting of the pkSZ and 21\,cm power spectra, resulting in a degeneracy between them. The pkSZ data seems to favour a low $\kappa$, which in order to match the measurement is then compensated by a low log$_{10}(\alpha_{0})$. Consequently, while the true pkSZ power spectra peaks at $\ell \approx 3000$, here there are multiple models for which the pkSZ reaches its maximum amplitude at $\ell \approx 2000$. Part of this bias stems from the simplifications in equation~\eqref{eq:recons_21cm_ps_simplified}, which does not fully account for the cross-correlation between the ionisation and density fields (see section~\ref{subsec:3_recons}) and results in a boosted amplitude of the 21\,cm power spectrum. As the \texttt{opkSZ} case does not contain information from 21\,cm data points, these models are not constrained, resulting in a biased distribution. We confirm that this effect does not occur in the \texttt{o21} case, where we only use the 21\,cm data points. However, as both data points are for the same k value, the parameter values are not well constrained (see section~\ref{subsec::2k1z}). 

%We have chosen these data points as our main model to investigate the amount of information we can extract about reionization, with the fewest data points. The advantage of minimalist approach is twofold. On the one hand, we model the constrain power of early stages of SKA-\textit{Low} observations on the epoch of reionization. On the other hand, we can better explore the role of each data point on our forecast as well as limitations. 
To summarise, we find that with as few as 100\,hours of integration with SKA-Low, we can successfully constrain the global reionization parameters $z_{\mathrm{re}}$ and $z_{\mathrm{end}}$, the data at different redshifts giving access to the evolution of the 21\,cm signal. %We confirm this by examining the role each parameter has on our observables. For example, the first two columns of Fig.~\ref{fig:app_params_vs_data} directly illustrate the role of the global history parameters on the ionisation history as well as the amplitude of the pkSZ power spectrum. 
By conducting our analysis with either only the pkSZ or only the 21\,cm data, we find that, while the 21\,cm signal can limit the range of values of $z_{\mathrm{end}}$, it cannot constrain $z_{\mathrm{re}}$. Conversely, we find that the pkSZ measurement mainly constrains the duration of reionization (d$z = z_{\mathrm{re}} - z_{\mathrm{end}}$). Naturally, the combined analysis benefits from the complimentarity of probes. While 100\,hours of integration are sufficient to obtain upper limits on the reionization morphology parameters, it is only with 1000\,hours that we are capable of constraining them. The morphology parameters have a direct impact on the shape of the 21\,cm power spectrum across different $k$-scales and we discuss how we can further constrain them in the following section.

\section{Discussion \& Limitations}
\label{sec:discussion}

No estimator is without its limitations and it is essential to consider its biases to better grasp the scope of the results. In this section, we delve into a series of cases designed to examine the capabilities of the forecast. We vary the redshifts and $k$-ranges at which the data points were selected in section~\ref{sec::SKA_ideal}, as well as the noise models and study the impact on our results. We also explore how the choice of input parameters affects the forecast. Note that for all cases in this discussion, we limit ourselves to the fixed number of data points as in section~\ref{sec::SKA_ideal} to quantify the role of each data point and its driving physical processes. The interested reader can refer to section~\ref{sec::3k2z} for results on how an additional 21\,cm data point can significantly improve the forecast.

\subsection{Importance of the choice of data points}
\label{sec:importance_of_data_points}

\subsubsection{Choice of redshift}
\begin{figure}
    \centering
    \includegraphics[width=\columnwidth]{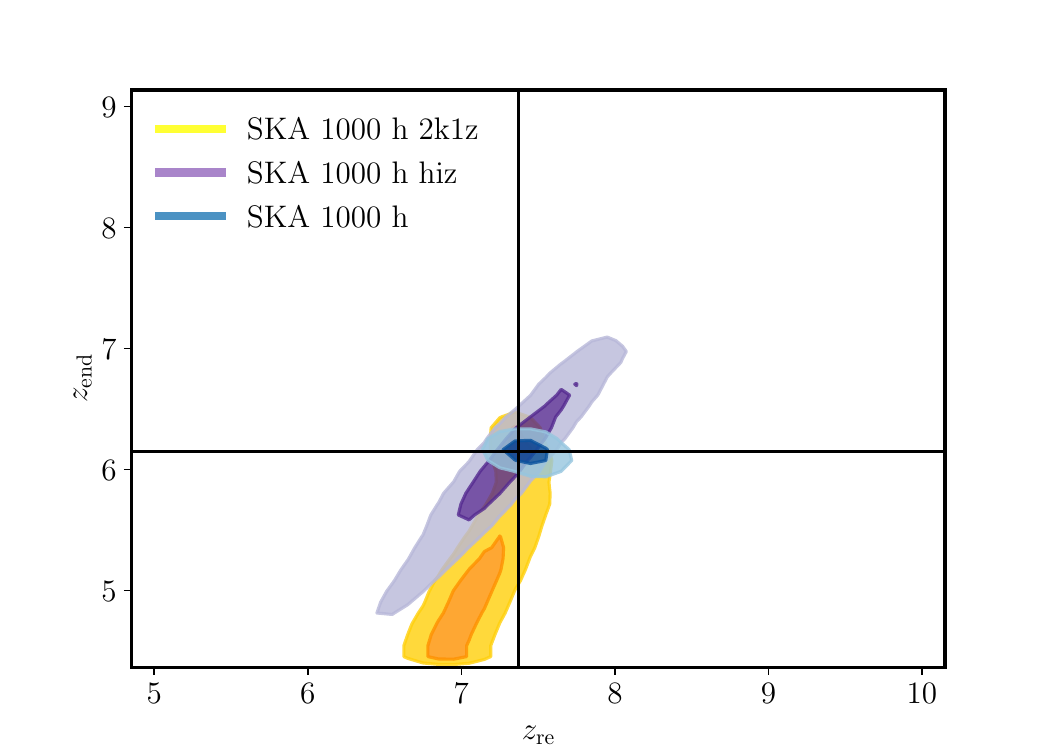}
    \includegraphics[width=\columnwidth]{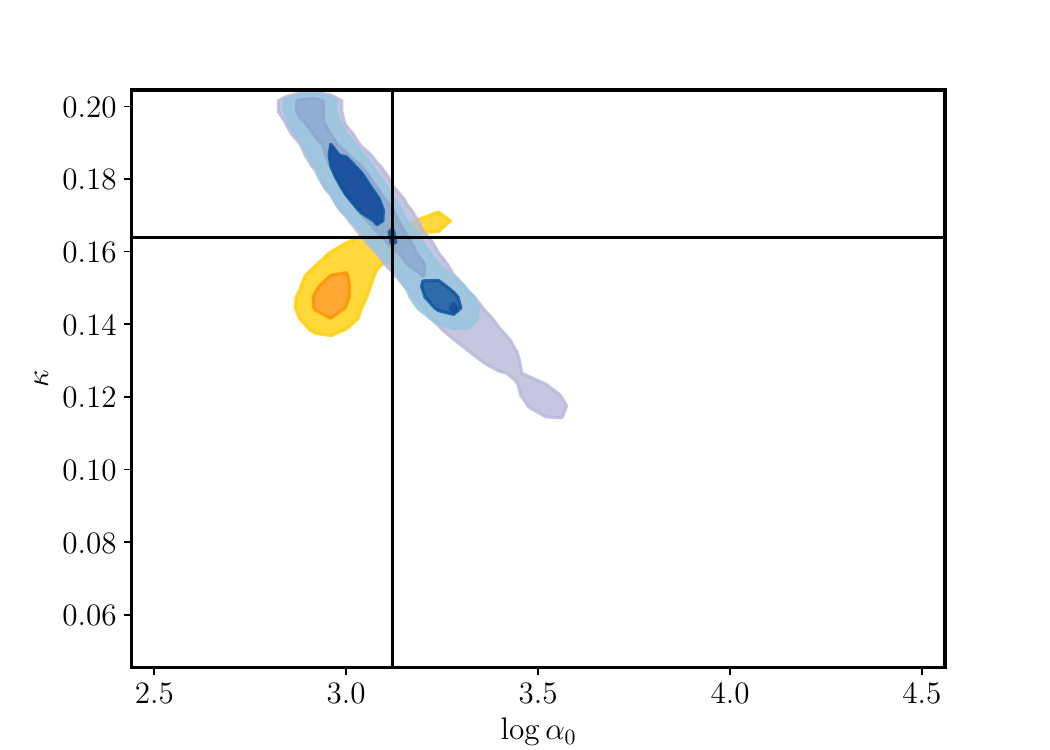}
    \caption{Posterior distributions on reionization mid- and endpoint (upper panel) and the morphology parameters log$_{10}(\alpha_{0})$ and $\kappa$ (bottom panel) using 21\,cm power spectra measurements from redshifts 7.8 and 10.4 at k = 0.5~Mpc$^{-1}$ and kSZ power spectrum measurement at $\ell =3000$ for 1000~hours of SKA-\textit{Low} observations compared to the fiducial case from Sec.~\ref{subsec:3_fit} (in blue). The vertical and horizontal lines show the `true' values used to generate the mock data.}
    \label{fig:2k1z_zre_zend_comparison}
\end{figure}

We explore the constraining ability of the forecast based on the choice of redshifts for the 21\,cm observation. To do so, we examine the SKA-\textit{Low} 1000h case (\texttt{hiz}) and vary the redshift of the observed 21\,cm power spectrum from $z = 6.5$ and 7.8 to $z = 7.8$ and 10.4, motivated by the recent HERA upper limits \citep{HERA2021}. The resulting two-dimensional posterior distributions for $z_\mathrm{re}-z_\mathrm{end}$ and $\log \alpha_0-\kappa$ are presented in Fig.~\ref{fig:2k1z_zre_zend_comparison} in purple. Compared to the fiducial case (in blue) two main differences are apparent. First, the data are now compatible with models whose mid- and endpoint are generally at higher redshifts (early and rapid reionization scenarios). For comparison, for the \texttt{1k2z} cases in section~\ref{sec::SKA_ideal}, the fact that the 21\,cm signal was non-zero at $z = 6.5$ naturally excludes models where reionization was completed by then. Additionally, the high-redshift data allow for later reionization histories than the truth (lower left quadrant of the figure). Indeed, the data point at $z = 10.4$ has limited constraining power on the global history parameters but provides better constraints on morphology parameters, as illustrated in the bottom panel of Fig.~\ref{fig:2k1z_zre_zend_comparison}, in orange. The reason for this is that at that redshift, $x_v \ll 1$, such that $P_{21} \sim P_{bb}+x{_m}^{2} P_{ee}$ and we do not vary the baryon density power spectrum in our analysis. Meanwhile, the shape of the electron density power spectrum is governed by the power law in equation~\eqref{eq:Pee_model}. 

\subsubsection{Choice of $k$-scales}
\label{subsec::2k1z}

\begin{figure}
    \centering
    \includegraphics[width=1.\columnwidth]{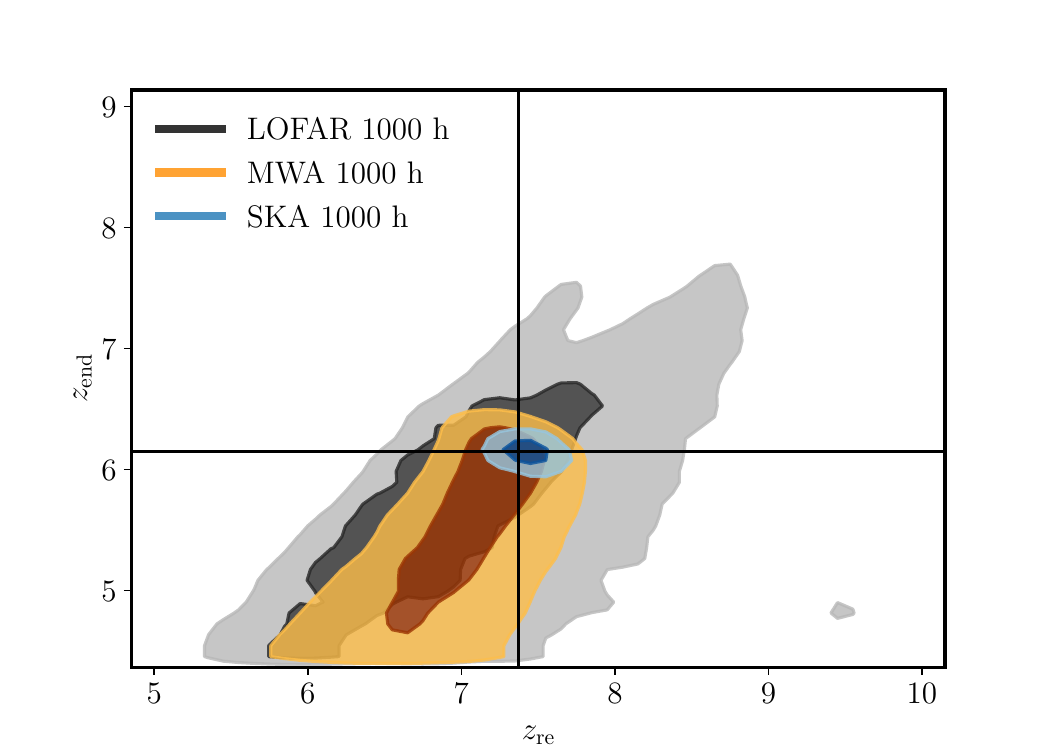}
    \includegraphics[width=1.\columnwidth]{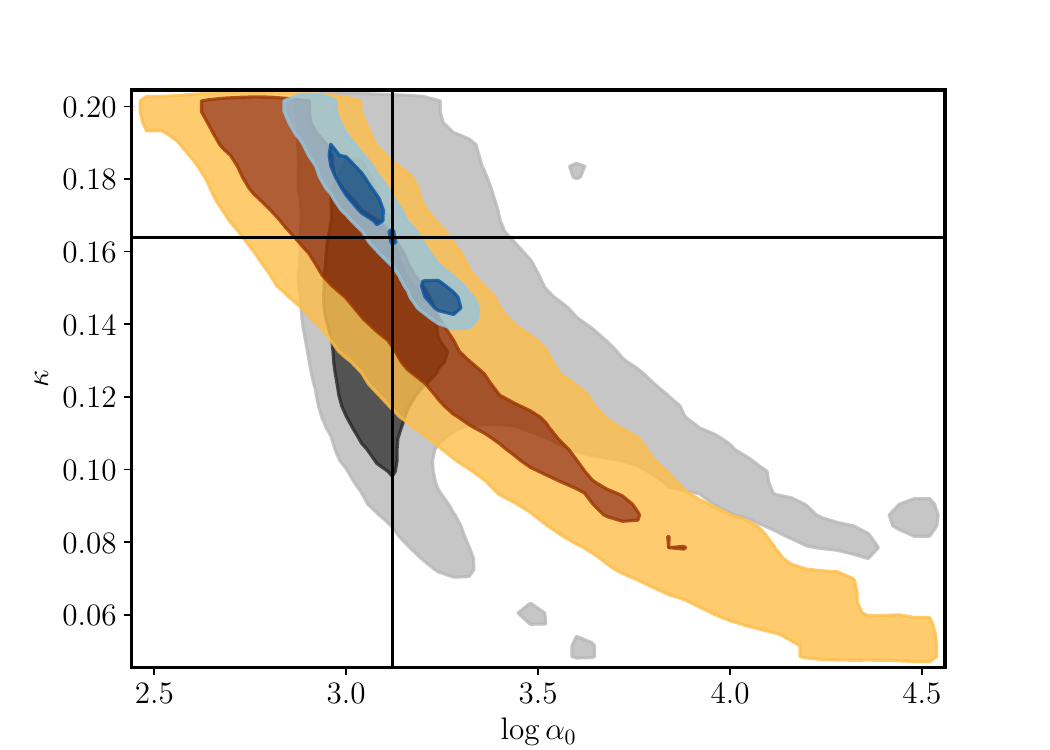}
    \caption{ Posterior distributions on reionization mid- and endpoint (upper panel) and the morphology parameters log$_{10}(\alpha_{0})$ and $\kappa$ (bottom panel) when fitting 21\,cm and pkSZ power spectra jointly, for 1000~hours of MWA, LOFAR, \& SKA observations, seen in grey, orange, and blue. The vertical and horizontal lines show the `true' values used to generate the mock data.}
    \label{fig::triangle_lofar_mwa}
\end{figure}

\begin{table*}
    \centering
    \begin{tabular}{lcccccccc}
        Models & & & $z_{\mathrm{re}}$ & $z_{\mathrm{end}}$ & log$_{10}(\alpha_{0})$ & $\kappa$ & $\tau$ & dz  \\
        \hline\hline
        Label & Data &True & 7.37  & 6.15 &  3.12 & 0.16 & 0.0649 & 1.22  \\
        \hline
        \texttt{MWA}& z = 6.5, 7.8 & 1000h & 7.02 $\pm$ 0.49 & 5.39 $\pm$ 0.54  & 3.25 $\pm$ 0.45 &  0.13 $\pm$ 0.04  & 0.0623 $\pm$ 0.0051  & 1.53 $\pm$ 0.49\\
        & k = 0.1\,Mpc$^{-1}$ &R - 1  & 0.01  & 0.008  & 0.03 & 0.03  & N/A  & N/A\ \\
        \hline 
        \texttt{LOFAR}& z = 8.3, 9.1 & 1000h  & 7.30 $\pm$ 0.87  & 5.74 $\pm$ 0.80  & 3.15 $\pm$ 0.33 &  0.14 $\pm$ 0.03 & 0.0643 $\pm$ 0.0094  & 1.38 $\pm$ 0.83\\
         & k = 0.1\,Mpc$^{-1}$ & R - 1  & 0.06  & 0.03  & 0.1 &  0.04  & N/A  & N/A\ \\
        \hline 
    \end{tabular}
    \caption{Best fit values of the distributions in Fig.~\ref{fig::triangle_lofar_mwa} and their corresponding 1$\sigma$ uncertainty as well as the Gelman–Rubin convergence diagnostic for each parameter.}
    \label{tab:LOFAR_MWA_MCMC_results}
\end{table*}

\begin{figure}
    \includegraphics[width=\columnwidth]{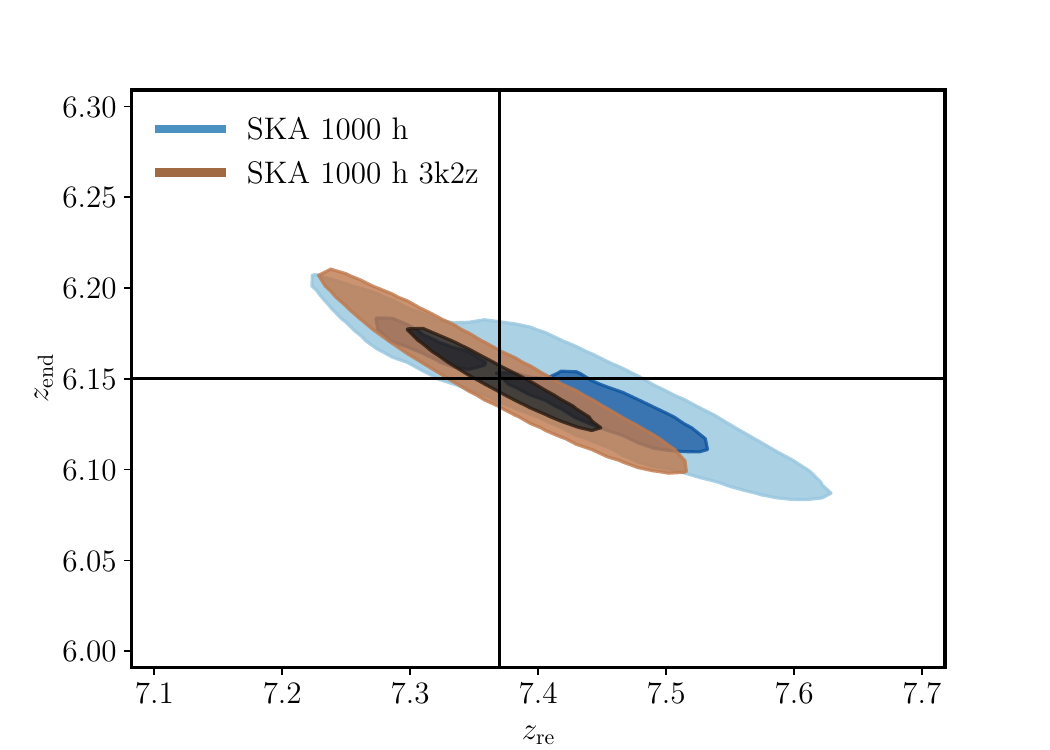}
    \includegraphics[width=\columnwidth]{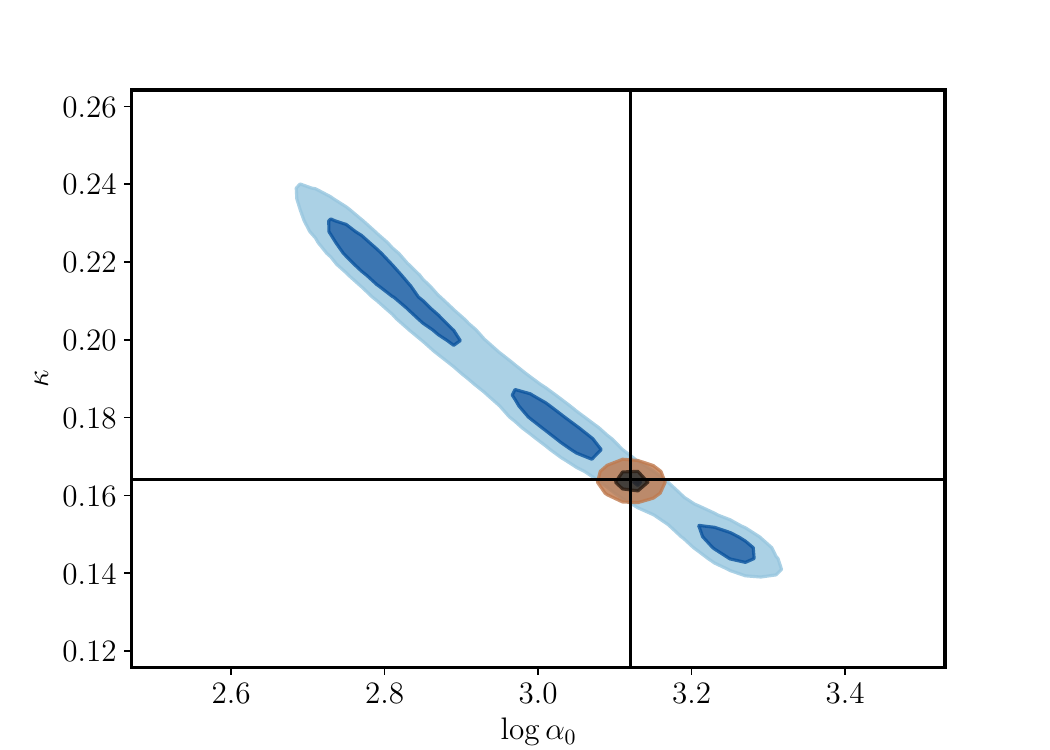}
    \caption{Posterior distributions on reionization mid- and endpoint (upper panel) and the morphology parameters log$_{10}(\alpha_{0})$ and $\kappa$ (bottom panel), considering a detection of the pKSZ power spectrum measurement at $\ell = 3000$ and of the 21\,cm power spectrum at redshifts of $z =6.5, 6.5, 7.8$ at $k = 0.1, 0.50, 0.5$\,Mpc$^{-1}$ for a 1000~hours of integration time with SKA, seen in brown and compared to the \texttt{1k2z} 1000h fiducial case shown in blue (see Fig,~\ref{fig::SKA_ideal}).}
    \label{fig::SKA_3k2z}
\end{figure}

\begin{table*}
    \centering
    \begin{tabular}{lcccccccc}
        Models & &  & $z_{\mathrm{re}}$ & $z_{\mathrm{end}}$ & log$_{10}(\alpha_{0})$ & $\kappa$ & $\tau$ & dz  \\
        \hline\hline
        Label & Data &True & 7.37  & 6.15 &  3.12 & 0.16 & 0.0649 & 1.22  \\
        \hline
        \texttt{2k1z}& z = 6.5, 6.5 & 1000h & 7.22 $\pm$ 0.21 & 5.29 $\pm$ 0.55  & 2.97 $\pm$ 0.08 &  0.15 $\pm$ 0.01  & 0.0649 $\pm$ 0.0016  & 1.90 $\pm$ 0.40\\
         & k = 0.1, 0.5\,Mpc$^{-1}$ & R - 1  & 0.3  & 0.9  & 4.0 & 1.2  & N/A  & N/A\ \\
        \hline 
        \texttt{hiz}& z = 7.8, 10.4 & 1000h  & 7.28 $\pm$ 0.33  & 6.02 $\pm$ 0.52  & 3.09 $\pm$ 0.15 &  0.17 $\pm$ 0.02 & 0.0642 $\pm$ 0.0025 & 1.27 $\pm$ 0.19\\
         & k = 0.1, 0.1\,Mpc$^{-1}$ & R - 1  & 0.7  & 0.7  & 0.5 & 0.5  & N/A  & N/A\ \\
        \hline 
        \texttt{3k2z}& z = 6.5, 6.5, 7.8 & 1000h  & 7.37 $\pm$ 0.07  & 6.15 $\pm$ 0.03  & 3.12 $\pm$ 0.01 &  0.164 $\pm$ 0.001 & 0.0649 $\pm$ 0.0010 & 1.22 $\pm$ 0.10\\
         & k = 0.1, 0.5, 0.1\,Mpc$^{-1}$ & R - 1  & 0.02  & 0.02  & 0.005 & 0.004  & N/A  & N/A\ \\
        \hline 
    \end{tabular}
    \caption{Best fit values of the distributions in and their corresponding 1$\sigma$ uncertainty as well as the Gelman–Rubin convergence diagnostic for each parameter.}
    \label{tab:otherMCMC_results}
\end{table*}

Next, we investigate the constraining power of the forecast using two detections of the 21\,cm power spectrum at $z = 6.5$ and $k =0.1$ and 0.5\,Mpc$^{-1}$. We assume thermal noise errors corresponding to 1000\,hours of observations with SKA-\textit{Low}. As before, a ten per cent error bar is assumed on the pkSZ power spectrum. We refer to this case as \texttt{2k1z}. Previously, we utilised the fact that reionization is an evolving process, whilst here, we make use of its multi-scale nature. Cosmic reionization is thought to be an inside-out phase transition, where the densest parts of the Universe are the first to be re-ionised and cosmic voids are the last \citep{Iliev2014}. This will impact the shape of the 21\,cm power spectrum, specifically, the shape of the high-$k$ `tail' which encodes information about the IGM on small scales. 

Fig.~\ref{fig:2k1z_zre_zend_comparison} illustrates the constraining power of this approach and results for \texttt{2k1z} are shown in yellow, in contrast to the fiducial \texttt{1k2z} case, shown in blue. While the data have clear constraining power on the $z_{\mathrm{re}}$ and $z_{\mathrm{end}}$ values, the two-dimensional posterior distribution exhibits a tail and is biased towards late reionization models ($z_{\mathrm{end}} = 5.29 \pm 0.55$, whilst the true value is at redshift 6.15). Observing a non-zero 21\,cm data point close to the true $z_\mathrm{end}$ naturally excludes early reionization models. Despite the increased measurement error, the low-$k$ (large-scale) amplitude of the 21\,cm power spectrum constrains the upper limit on the midpoint. Indeed, models with an early midpoint of reionization would either re-ionise earlier or have a lower large-scale power, which is inconsistent with the measurement. Additionally, as can be inferred from Fig.~\ref{fig:tel_noise}, the low-$k$ 21\,cm measurement has a higher uncertainty, primarily due to a larger sample variance. This results in a poorly constrained tail of the 21\,cm power spectrum, which undermines our ability to probe the state of the IGM with a single redshift measurement. %Fig.~\ref{fig:tel_noise} also indicates that lower $k$-scales are subject to higher noise levels. 
Compared to the fiducial \texttt{1k2z} case, the choice of probing two $k$-scales results in a tighter constraining power on log$_{10}(\alpha_{0})$ and $\kappa$, although the two-dimensional posterior is biased towards lower values of the parameters, corresponding to later reionization models. Examining and comparing the 21\,cm power spectra in Fig.~\ref{fig:app_params_vs_data} reveals that the differences in $k$-scale are most prominent for $k>1.0$ Mpc$^{-1}$ scales. Since the high $k$-scale measurements of the 21\,cm power spectrum exhibit a higher level of uncertainty due to larger instrumental noise (see Fig.~\ref{fig:tel_noise}), models are less tightly constrained on such scales. \\

In summary, with two 21\,cm data points at different $k$-scales, the morphology parameters are better constrained as the scale-evolution of the 21\,cm power spectrum is included in the forecast. This result is, however, mitigated by the fact that, %the \texttt{2k1z} case has a greater constraining power of the morphology parameters of reionization, but
by fixing the redshift, {the uncertainty on the middle and end of reionization is increased, resulting in a tail of the $z_{\mathrm{re}}$-$z_{\mathrm{end}}$ posterior as well as biased morphology parameters. One way to improve constraints on $z_\mathrm{end}$ would be to consider 21\,cm data for $k>1.0$ Mpc$^{-1}$ scales, or to limit the prior on $z_{\mathrm{end}}$ to values allowed by measurements of Ly~$\alpha$ absorption in quasar spectra \citep{Bosman2022}.} 

\subsection{Current Telescope Capabilities}
\label{sec::lofar_mwa}

Currently operating radio telescopes are getting closer and closer to a detection of the 21\,cm power spectrum during the EoR. In this section, we examine the constraining ability of the forecast, assuming a potential measurement of the 21\,cm power spectrum by the MWA and LOFAR radio telescopes, i.e., with higher noise levels than previously considered\footnote{Note that we do not consider HERA in this section, since its ideal noise levels are lower than the ones for the SKA (Fig.~\ref{fig:tel_noise}).}
Because of its specific characteristics, each telescope is sensitive to different redshifts and scales and, in turn, to different reionization parameters. For this reason, we choose LOFAR data points corresponding to a fixed $k$-scale of $k = 0.1$ Mpc$^{-1}$ for redshifts $z = 8.2$ and 9.1 \citep[based on][]{Patil2017,Mertens2020}. The MWA data is consistent with the data points chosen in section~\ref{sec::SKA_ideal}. For both cases, we assume an integration time of 1000\,hours. The resulting two-dimensional posterior probability distributions of the cosmological and morphology parameters are presented in Fig.~\ref{fig::triangle_lofar_mwa} and Table. \ref{tab:LOFAR_MWA_MCMC_results}. 

Results for the \texttt{MWA} case are overall consistent with section~\ref{sec::SKA_ideal}, with key differences emerging due to the higher uncertainty on the 21\,cm power spectrum, which is approximately 1000~times larger than that of the SKA (see Fig.~\ref{fig:tel_noise}). As shown in orange in the upper panel of Fig.~\ref{fig::triangle_lofar_mwa}, we are still able to constrain $z_{\mathrm{re}} = 7.02 \pm 0.49$. However, the spread in the distribution has increased compared to the SKA-\textit{Low} case in blue (for which $z_{\mathrm{re}} = 7.39 \pm 0.14$), and the result is biased to low values (see Table~\ref{tab:rsage_ref_params}). On the other hand, the constraint on the end of reionization is now limited to an upper limit $z_{\mathrm{end}} <7.5$. A lower constraining power is also observed for the morphology parameters. We can primarily obtain upper limits for log$_{10}(\alpha_{0}) < 3.75$ and $\kappa >0.08$, showcasing the ability of the mock \texttt{MWA} data to exclude outlying models of the EoR. The decrease in the overall constraining power of the forecast is linked to the increase in the uncertainty of the 21\,cm power spectrum points. This increase results in favouring late-time reionization models where the 21\,cm power spectrum at $z =6.5$ has a higher amplitude and a lower value of $\kappa$, and the pkSZ power spectrum peaks at $\ell \approx 2000$, similar to that discussed in section~\ref{secc:each_data_set} for the \texttt{opKSZ} case. Such models could potentially be excluded with the addition of a pkSZ data point at $\ell = 2000$ or more data on the 21\,cm power spectrum at lower-$k$ values.

Regarding the \texttt{LOFAR} case, results are roughly similar to what was observed in Sec.~\ref{subsec::2k1z} where the 21\,cm mock data is taken at a fixed $k$-scale but different redshifts. Naturally, increasing the uncertainty on the measurements by approximately two orders of magnitude compared to that of the SKA (see Fig.~\ref{fig:tel_noise}), the constraining power of the data is decreased. The choice of a higher-$z$ mock 21\,cm data point further reduces the ability of the forecast to place upper limits on the global reionization parameters (see Sec.~\ref{sec:importance_of_data_points}). For example, as seen in grey in the upper panel of Fig.~\ref{fig::triangle_lofar_mwa}, we can constrain $z_{\mathrm{re}} <8.0$ and can only exclude only late reionization models for which $z_{\mathrm{end}} > 5.5$. Compared to the \texttt{MWA} case, having higher-$z$ 21\,cm data leads to higher constraining power on the morphology parameters. While we can only place an upper limit on $\kappa >0.1$, the constraint on log$_{10}(\alpha_{0})  =3.15 \pm 0.33$ is fairly comparable to that of \texttt{1k2z} 1000\,hours (where log$_{10}(\alpha_{0})  = 3.04 \pm 0.32$). While this is mostly related to the choice of data, for example, the \texttt{hiz} case can constrain log$_{10}(\alpha_{0})= 3.09 \pm 0.15$, our results indicate the potential of high-redshift LOFAR measurements on constraining on the morphology parameters of the EoR.

A possible improvement is the addition of an informed prior on the Thomson optical depth from the \citet{PlanckCollaborationAghanim_2020} measurement, which could potentially aid in increasing the precision of the forecast and constraining the parameter space of the global history parameters. %Note, however, that a misinformed prior would most likely result in the opposite and will lead to a biased result.

\subsection{Increasing the number of data points}
\label{sec::3k2z}

In previous sections, we used a minimalist approach by limiting the number of detected modes to test how much information would could obtain from early, partial 21\,cm data, as a detection will only be achieved incrementally. On the one hand, with with two data points at a fixed $k$-scale ($k = 0.5$\,Mpc$^{-1}$ for $z = 6.5,7.8$, \texttt{1k2z} case), the forecast is sensitive to the redshift evolution of the 21\,cm signal and we are able to well constrain the global reionization parameters $z_{\mathrm{re}}$ and $z_{\mathrm{end}}$. On the other hand, with two data points at fixed redshift but different scales ($z = 6.5$ for $k = 0.1,0.5$\,Mpc$^{-1}$, \texttt{2k1z} case), we can constrain the morphology parameters log$_{10}(\alpha_{0})$ and $\kappa$. A natural way to extend these results is to venture into the multi-scale regime (\texttt{3k2z} case, seen in brown in Fig.~\ref{fig::SKA_3k2z}) by combining both of the aforementioned data sets such that the 21\,cm power spectrum is measured at $z=6,5,6.5,7.8$ and $k = 0.1,0.5,0.5$ Mpc$^{-1}$ for 1000\,hours of integration with the SKA-\textit{Low}.

%We examine the results, noting that the limits of the axes differ from the previously presented ones.
We find that the forecast inherits the merits of the \texttt{1k2z} and \texttt{2k1z} cases as we access information on both the amplitude and the shape of the 21\,cm power spectrum.
Compared to our primary \texttt{1k2z} case (seen in blue in Fig.~\ref{fig::SKA_3k2z}), we improve the accuracy of our constraints of the reionization parameters: We get $z_{\mathrm{re}} = 7.37 \pm 0.07$ and $z_{\mathrm{end}} = 6.15 \pm 0.03$, significantly reducing the 1$\sigma$ standard deviation (especially for $z_{\mathrm{end}}$). 
Consequently, we can tightly constrain the parameter space around the true value and retrieve the correct Thomson optical depth with a precision of $\Delta \tau = \pm 0.001$ (see Appendix~\ref{appendix::tau}). Hence, our method could provide an independent measurement of the optical depth, separate from the analysis of large-scale CMB polarisation anisotropies \citep{Planck2016_XLVII}, and potentially break well-known degeneracies with other cosmological parameters \citep{Liu2016}. Conversely, large-scale CMB data could be included in the analysis in order to improve constraints on the reionization parameters \citep{GorceDouspis_2022}. 

Constraints on the morphology parameters are significantly tighter: We measure log$_{10}(\alpha_{0}) = 3.12 \pm 0.01$ and $\kappa = 0.164 \pm 0.001$. This improvement is partly explained by the fact that adding data at small scales breaks the degeneracy between the morphology parameters. We can intuitively explain this by referring to the results in section~\ref{subsec::2k1z}: The \texttt{3k2z} case can be understood as the combination of  the \texttt{1k2z} and \texttt{2k1z} cases and we see on Fig.~\ref{fig:2k1z_zre_zend_comparison} that accessing the redshift-evolution of the 21\,cm signal with data points at different redshifts removes the bias on the morphology parameters obtained with \texttt{2k1z}. 

Looking at these results, it is tempting to think that all the constraining power comes from the three 21\,cm data points. However, we have also examined an additional case where we only consider the 21\,cm mock data from \texttt{3k2z} and exclude the pkSZ data point. We find that, while we can recover the morphology parameters with similar, albeit weaker, constraints, the global history parameters are significantly more degenerate. Indeed, the information the pkSZ data point provides on the duration of reionization remains crucial for the accuracy of the forecast. Measuring the pkSZ spectrum at different multipoles would enable tighter constraints on the shape parameter $\kappa$, whether at lower ($\ell=2000$) or larger ($\ell=5000$) multipoles. However, note that the former will be more challenging to achieve observationally because of the extremely large amplitude of the primary CMB temperature power spectrum on scales $\ell \lesssim 2000$.

If we have previously kept our results to simplistic cases in order to provide a proof-of-concept, these further tests show the full potential of our approach: As more and better measurements of the pkSZ and the 21\,cm power spectra become available, this method will not only enable constraints on the global history of reionization but also on its morphology.

\section{Conclusions}
\label{sec:conclusions}

In this work, we examine how the fundamental relation between the patchy kSZ effect and the 21\,cm signal can help us constrain the nature of cosmic reionization. In section~\ref{sec:methods}, we express and relate the 21\,cm and kSZ power spectra through the parameterisation of the electron power spectrum $P_{ee}(k,z)$ presented in \citet{GorceIlic_2020}. The resulting equation~\eqref{eq:recons_21cm_ps_full}, while linking both observables, contains non-trivial cross-correlation and second-order terms of the ionised and density fields. We choose to ignore such terms in this work, leading to equation~\eqref{eq:recons_21cm_ps_fit}. Looking at the semi-numerical {\sc RSAGE} simulation \citep{Seiler2019}, we find that this assumption leads to overestimating the 21\,cm power on all scales and redshifts $z \lesssim 10$. 

Aware of this limitation, we use the derived relation in a forecast analysis using an MCMC sampling method: We fit for the reionization mid- and endpoint, $z_\mathrm{re}$ and $z_\mathrm{end}$, as well as for two reionization morphology parameters log$_{10}(\alpha_{0})$ and $\kappa$. Indeed, for each of set of these parameters, we can derive the electron power spectrum and, in turn, the kSZ and 21\,cm power spectra. We generate mock observations of these with the parameter values given in Table~\ref{tab:rsage_ref_params}, assume a ten per cent error bar for the pkSZ data point at $\ell = 3000$, %of the global reionization parameters are $z_{\mathrm{re}} = 7.37$, $z_{\mathrm{end}} = 6.15$, while for the morphological parameters log$_{10}(\alpha_{0}) = 3.12$, and $\kappa = 0.16$ (see Table~\ref{tab:rsage_ref_params}).
and include thermal noise and sample variance in the 21\,cm measurement errors, typically for 1000\,hours of integration with SKA-\textit{Low}.
We follow a minimalist approach as we do not assume a full detection of the kSZ and 21\,cm power spectra over a range of scales, but assess the constraining power of only a few observed data points. Such an approach gives us insight on how much we can learn about reionization with early measurements of both observables. Our findings are as follows.  \\

With as few as 100\,hours of SKA-\textit{Low} observations and two 21\,cm measurements at $k = 0.5$\,Mpc$^{-1}$ and $z = 6.5, 7.8$ (referred to as the \texttt{1k2z} case, shown in blue in Fig.~\ref{fig::SKA_ideal} and throughout this work), we are able to constrain the global reionization parameters $z_{\mathrm{re}} = 7.42 \pm 0.13$ and $z_{\mathrm{end}} = 6.15 \pm 0.04$ and provide limits on the morphology parameters log$_{10}(\alpha_{0}) > 2.64$ and $\kappa < 0.25$ (Table~\ref{tab:MWA_MCMC_results}). We demonstrate that this constraining power stems from the complementary nature of both data sets: The 21\,cm signal is mostly sensitive to the endpoint of reionization, while the pkSZ effect is inherently sensitive to its duration d$z=z_\mathrm{re}-z_\mathrm{end}$. Our results show that, even the early stages of SKA-\textit{Low} operations, its observations will provide valuable insight into the global history of the EoR despite a minimal knowledge of the properties of the first stars and galaxies. The constraints on the reionization global history are further improved when the fitted mock 21\,cm measurements are at different redshifts (but a given $k$-scale), as the data now trace the redshift-evolution of the 21\,cm signal. 

We show that the choice of redshift for the mock data set plays an important role. First, any low-$z$ measurement excludes earlier reionization models, for which the 21\,cm signal would be zero.
Second, at high redshift, when the global ionisation fraction is $x_v \ll 1$, the data will primarily determined by log$_{10}(\alpha_{0})$ and $\kappa$. In our tests, for 21\,cm mock measurements at high(er) redshift (\texttt{hiz} case, $z = 7.8$ and 10.4 at $k = 0.5$\,Mpc$^{-1}$), the constraints on the global reionization parameters $z_{\mathrm{re}}$ and $z_{\mathrm{end}}$ are loosened compared to the \texttt{1k2z} case, at $z=6.5$ and 7.8 (Fig.~\ref{fig:2k1z_zre_zend_comparison}). On the other hand, it is precisely the constraining of the high-$z$ 21\,cm power spectrum which allows for firmer constraints on the morphological parameters: log$_{10}(\alpha_{0}) = 3.10 \pm 0.17$ and $\kappa = 0.17 \pm 0.02$.

We also observe that the $k$-scale of the data significantly influences forecast constraints. Indeed, the inside-out nature of cosmic reionization will impact the shape of the 21\,cm power spectrum, specifically, its high-$k$ `tail', which encodes information about reionization morphology on small scales. % Informed by Fig.~\ref{fig:tel_noise}, we choose 21\,cm power spectrum data at $k = 0.1, 0.5$ Mpc$^{-1}$ to balance out the uncertainty between sample variance and instrumental noise.} 
By choosing two mock 21\,cm power spectra at a fixed redshift (\texttt{2k1z}, $z = 6.5$  at $k = 0.1, 0.5$ Mpc$^{-1}$), we find that the constraints on the global reionization parameters are weakened to upper limits (Fig.~\ref{fig:2k1z_zre_zend_comparison}, in yellow) but constraints on the morphology parameters are improved. However, the recovered values are biased by 6\%. We find that major differences between the allowed models (which reach the mid- and endpoint later than the true model) and the true 21\,cm power spectrum are most distinguishable for modes $k > 1.0$ Mpc$^{-1}$, for which the model does not possess information. This implies that while multi-scale observations of the 21\,cm power spectrum at one redshift contain information on the whole process of reionization, it is the highest-$k$ modes, which are the most sensitive. However, such measurements are non-trivial because the measurement noise scales with $k$-scale (see Fig~\ref{fig:tel_noise}), making them less appealing in the context of early SKA-\textit{Low} observations.

It is also worth noting that we limit our main results to simplistic cases with two 21\,cm power spectrum observations in order to provide a proof-of-concept approach and to better showcase the workings of the forecast. Further tests where we increase the number of data points from two to three (e.g., the \texttt{3k2z} case seen in section~\ref{sec::3k2z}) show the full potential of this approach: Highlighted in brown in Fig.~\ref{fig::SKA_3k2z}, we see that as more and better measurements of the pkSZ and the 21\,cm power spectrum become available, this method will not only enable constraints on the global history of reionization but also its morphology. We obtain accurate constrains of the reionization parameters $z_{\mathrm{re}} = 7.37 \pm 0.07$ \& $z_{\mathrm{end}} = 6.15 \pm 0.03$ and the morphology parameters log$_{10}(\alpha_{0}) = 3.12 \pm 0.01$ \& $\kappa = 0.164 \pm 0.001$, significantly reducing the 1$\sigma$ standard deviation, compared to \texttt{1k2z}.

Finally, we explore forecast performance on mock data from operating telescopes such as MWA and LOFAR, based current upper limits. \citep{Trott2020,Patil2017,Mertens2020}. %For both data sets, we assume a measurement error associated with 1000\,hours of integration time and we base the choice of the mock data on 
Naturally, we find the mock data less constraining than in the case of \texttt{1k2z} (see Fig.~\ref{fig::triangle_lofar_mwa} in gray and orange) as can be expected from the higher uncertainty on the measurements. However, detections by both telescopes can place firm upper limits on the midpoint of reionization $z_{\mathrm{re}} \leq 8$. Moreover, the low-redshift observations of MWA (for the same set-up as \texttt{1k2z}) can also place a lower limit on the end of reionization $z_{\mathrm{end}} \geq 6.5$. Meanwhile, data from \texttt{LOFAR} ( chosen at $k = 0.1$ Mpc$^{-1}$ for $z = 8.3, 9.1$) is more sensitive to the morphology parameters, notably constraining log$_{10}(\alpha_{0}) = 3.15 \pm 0.33$. This implies that even before the first-light of SKA-\textit{Low}, the ongoing improvements of current 21\,cm power spectrum upper limits \citep[see fig. 6 of][for an example]{Raste2021} are likely to soon begin constraining the properties of reionization.\\

The main limitation of our model is the omission of the higher-order and cross-correlation terms in our simplified expression of the 21\,cm power spectrum in equation~\eqref{eq:recons_21cm_ps_fit}. Potential future developments would be develop analytic models of the cross-power spectrum and include them in the derivation \citep{McQuinn2005,Schneider2023}. However, this approach would most likely not capture the full complexity present at all $k$-scales and could be model-dependent. Another option would be to model the higher-order and the cross terms as constant biases and account for them as nuisances parameters in the forecast (see the discussion in appendix~\ref{appendix::bias}). Lastly, we plan to conduct our analysis for 21\,cm data over broader ranges of redshift and $k$-scales to explore the potential of joint analysis to identify and remove systematics such as residual foregrounds (see \citet{BeginLiu_2022} for such an analysis using the global 21\,cm signal).

\section*{Acknowledgements}

The authors thank Adrian Liu, Jack Line, Leon Koopmans, and Cathryn Trott for useful discussions about different parts of this work. We thank the anonymous reviwer for insightful and constructive
comments. IG acknowledges the support form the Gustaf och Ellen Kobbs stipendiestiftelse and the Alva and Lennart Dahlmark Research Grant.
At the time when most of this work was conducted, AG's work was supported by the McGill Astrophysics Fellowship funded by the Trottier Chair in Astrophysics, as well as the Canadian Institute for Advanced Research (CIFAR) Azrieli Global Scholars program and the Canada 150 Programme. 
GM’s research is supported by the Swedish Research Council project grant 2020-04691\_VR.

This research made use of \texttt{matplotlib}, a Python library for publication-quality graphics \citep{hunter_2007}, of \texttt{scipy}, a Python-based ecosystem of open-source software for mathematics, science, and engineering \citep{scipy} -- including \texttt{numpy} \citep{numpy}. The MCMC was performed with the \texttt{emcee} sampler \citep{emcee} and posterior distributions were drawn thanks to a modified version of the \texttt{corner} Python package \citep{corner}.
This research was enabled in part by support provided by Calcul Québec and the Digital Research Alliance of Canada\footnote{See \url{http://www.alliancecan.ca} and \url{https://www.calculquebec.ca}.}. Computations were notably performed on the Cedar supercomputer, hosted by the Simon Fraser University.

%%%%%%%%%%%%%%%%%%%%%%%%%%%%%%%%%%%%%%%%%%%%%%%%%%
\section*{Data Availability}

%The inclusion of a Data Availability Statement is a requirement for articles published in MNRAS. Data Availability Statements provide a standardised format for readers to understand the availability of data underlying the research results described in the article. The statement may refer to original data generated in the course of the study or to third-party data analysed in the article. The statement should describe and provide means of access, where possible, by linking to the data or providing the required accession numbers for the relevant databases or DOIs.

The sampling code used to obtain all the results presented in this article is publicly available at \url{https://github.com/adeliegorce/forecast_kszx21}. The (trained) random forests used to generate the angular patchy kSZ power spectra given a set of cosmological parameters can be found at \url{https://szdb.osups.universite-paris-saclay.fr}. Any additional information or data can be requested from the authors.

%%%%%%%%%%%%%%%%%%%% REFERENCES %%%%%%%%%%%%%%%%%%

% The best way to enter references is to use BibTeX:

\bibliographystyle{mnras}
\bibliography{biblio} % if your bibtex file is called example.bib

\appendix

% Alternatively you could enter them by hand, like this:
% This method is tedious and prone to error if you have lots of references
%\begin{thebibliography}{99}
%\bibitem[\protect\citeauthoryear{Author}{2012}]{Author2012}
%Author A.~N., 2013, Journal of Improbable Astronomy, 1, 1
%\bibitem[\protect\citeauthoryear{Others}{2013}]{Others2013}
%Others S., 2012, Journal of Interesting Stuff, 17, 198
%\end{thebibliography}

\section{Reconstructing the pkSZ from a measurement of the 21 cm power spectrum}
\label{app:obs_reconstruction}

\begin{figure}
    \centering
    \includegraphics[width=\columnwidth]{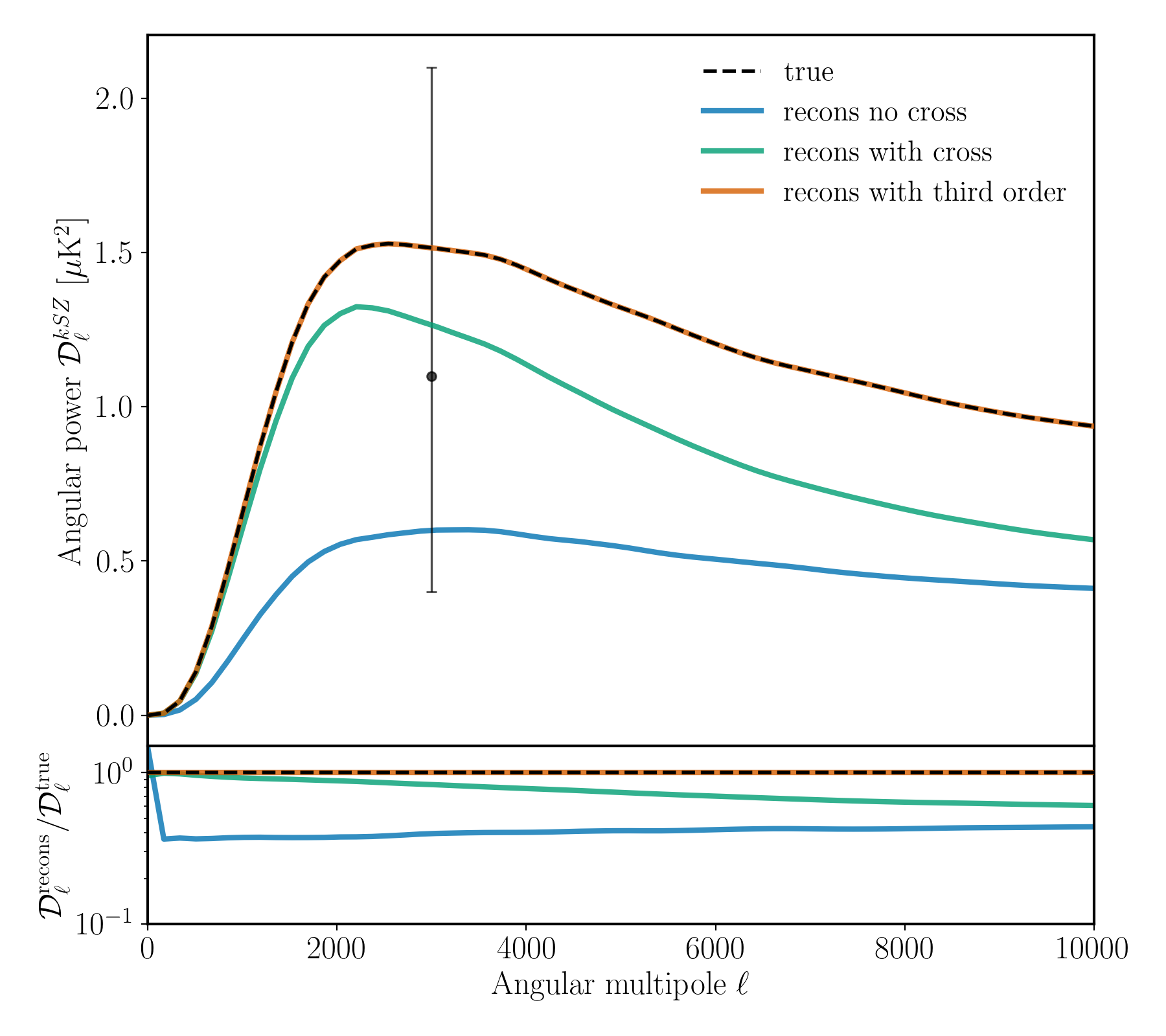}
    \caption{The pkSZ angular power spectrum obtained for the true electron power spectrum measured in the \texttt{RSAGE const} simulation, and for reconstructions based on the 21\,cm power spectrum at various precision levels. The colour scheme is similar to Fig.~\ref{fig:recons_comp}. The results are compared to the SPT data point on the pkSZ power \citep{ReichardtPatil_2021}. }
    \label{fig:pkSZ_recons_comp}
\end{figure}

Aware of the limitations of our reconstruction model presented in section~\ref{subsec:3_recons}, we now attempt to use the simplified 21\,cm power spectrum model to reconstruct the pkSZ angular power spectrum. That is, we compute numerically the 21\,cm power spectrum from our simulation and then use it to reconstruct the electron power spectrum following equation~\eqref{eq:recons_21cm_ps_full}. We then plug the reconstructed $P_{ee}(k,z)$ in equation~\eqref{eq:def_C_ell_pkSZ} to obtain the patchy kSZ angular power.
We compare results for different levels of reconstruction precision in Fig.~\ref{fig:pkSZ_recons_comp}, similarly to Fig.~\ref{fig:recons_comp}. Note that this figure is only intended as a comparison of the potential of the reconstruction, not as a precise estimate of the pkSZ power of the simulation\footnote{Remember that all cosmological and reionization parameters are derived from the \texttt{RSAGE} simulation.}. Indeed, the power is set to zero on $k$-modes not covered by the simulation. The results are slightly different from the ones presented in Fig.~\ref{fig:recons_comp}. Here, the pkSZ power is systematically under-estimated, as long as all the terms of equation~\eqref{eq:recons_21cm_ps_full} are not included. Despite our approximation largely underestimating the pkSZ power, the reconstructed power still lies within current error bars on pkSZ power spectrum measurements at $\ell=3000$ \citep{ReichardtPatil_2021}. 

\begin{figure}
    \centering
    \includegraphics[width=\columnwidth]{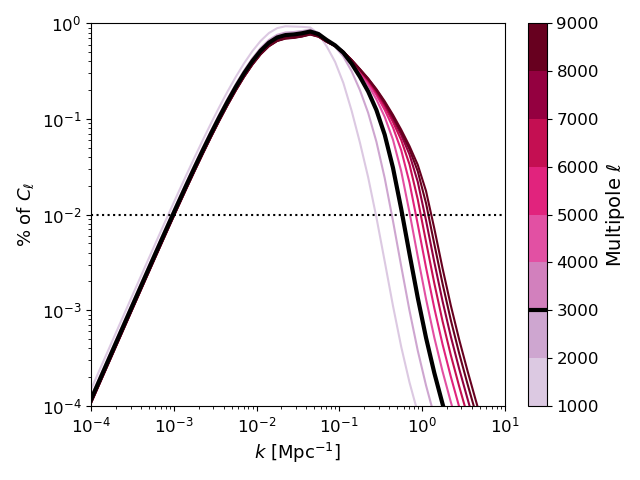}
    \caption{Physical scales contributing to the final pkSZ power at different angular multipoles $\ell$. The result for $\ell=3000$ is highlighted in bold.}
    \label{fig:contributions_to_pkSZ}
\end{figure}

Instead of reconstructing $P_{21}$ from equation~\eqref{eq:recons_21cm_ps_simplified} and missing out on crucial power, one could use direct measurements of the 21\,cm power spectrum to reconstruct the pkSZ power and compare to CMB observations. 
To do so, the former must be integrated over a wide range of scales. However, for now, the 21\,cm power spectrum has not been measured with sufficient precision on a wide enough range. We must therefore assess which scales contribute the most to the final pkSZ power and are necessary for this reconstruction. The relative contribution of each $k$-mode to the pkSZ $C_\ell$ at $\ell=3000$ had already been estimated in \citet{GorceIlic_2020}, but here we extend this result to the range $1000 \leq \ell  < 10\, 000$ and show the result in Fig.~\ref{fig:contributions_to_pkSZ}. We see that, for all multipoles, most of the power stems from $10^{-3} < k/[\mathrm{Mpc}^{-1}] < 1$ with only the upper limit increasing as $\ell$ increases.

\begin{figure}
    \centering
    \includegraphics[width=\columnwidth]{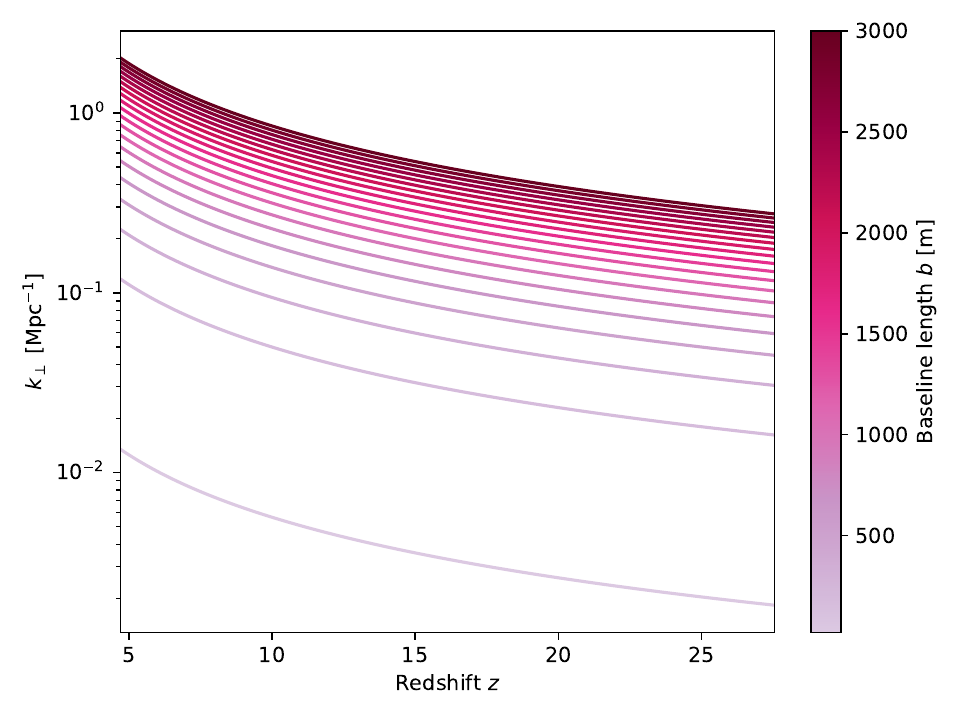}
    \includegraphics[width=\columnwidth]{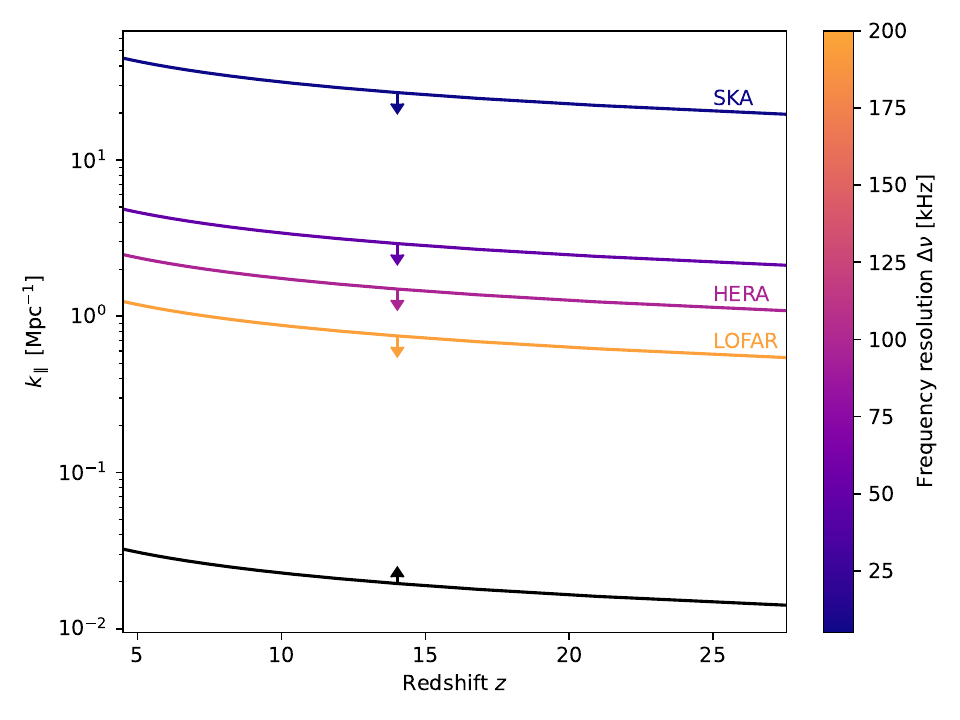}
    \caption{Longitudinal (lower panel) and transversal (upper panel) $k$-modes observable by an interferometer, depending on its characteristics such as baseline length $b$ or frequency resolution. The bandwidth is fixed to $15~\mathrm{MHz}$. In the lower panel, the frequency resolutions of SKA-\textit{Low}, HERA and LOFAR are shown as reference and result in an upper limit on the $k_\parallel$ modes that can be observed by these instruments.}
    \label{fig:reachable_k}
\end{figure}

The spherical Fourier modes used in Fig.~\ref{fig:contributions_to_pkSZ} are effectively the sum of a transversal and a longitudinal mode $k = \sqrt{k_\perp^2 + k_\parallel^2}$.
The Fourier modes probed by an interferometer measuring signal at a frequency $\nu$ corresponding to a redshift $z=\nu_{21}/\nu-1$ depend on its characteristics. The maximum and minimal $k_\perp$ probed will be a function of the bandwidth (of $\nu$) and of the baseline length $b$: 
\begin{equation} \label{eq:kperp}
k_\perp =  \frac{2\pi}{d_c(z)} \frac{\nu b}{c},
\end{equation}
where $d_c(z)$ is the comoving distance at redshift $z$.
On the other hand, the maximum and minimal $k_\parallel$ probed will depend on the bandwidth $B$ and the frequency resolution $\Delta \nu$ of the instrument:
\begin{equation} \label{eq:wf_centres}
\left\{ \begin{aligned}
&k_{\parallel, \mathrm{min}} =  \frac{2\pi}{\alpha(z)} \times \frac{1}{B},  \\
&k_{\parallel, \mathrm{max}} = \frac{2\pi }{\alpha(z)}\times \frac{1}{2 \Delta \nu} .
\end{aligned}\right.
\end{equation}
where $\alpha(z) \equiv c(1+z)^2/[\nu_{21}H(z)]$. These limits are presented in Fig.~\ref{fig:reachable_k} for various instruments, including SKA-\textit{Low}, HERA and LOFAR. In particular, for HERA, including the foreground wedge and additional buffer, we find that the accessible Fourier modes are limited to $0.15 \leq k / [\mathrm{Mpc}^{-1}] \leq 2.05$. Because of the foregrounds, the instrument cannot reach the large-scale modes required to estimate the pkSZ power spectrum.

\begin{comment}
\begin{figure}
    \centering
    \includegraphics[width=\columnwidth]{Figures/figA4.pdf}
    \caption{Same as Fig.~\ref{fig:recons_comp} but for k = 0.5 Mpc$^{-1}$, which is used in the analysis in section~\ref{subsec::2k1z}.}
    \label{fig:recons_comp_extra}
\end{figure}
\end{comment}

\section{Full posterior distributions}
\label{appendix::3k2z}

\begin{figure*}
    \includegraphics[width=1.\linewidth]{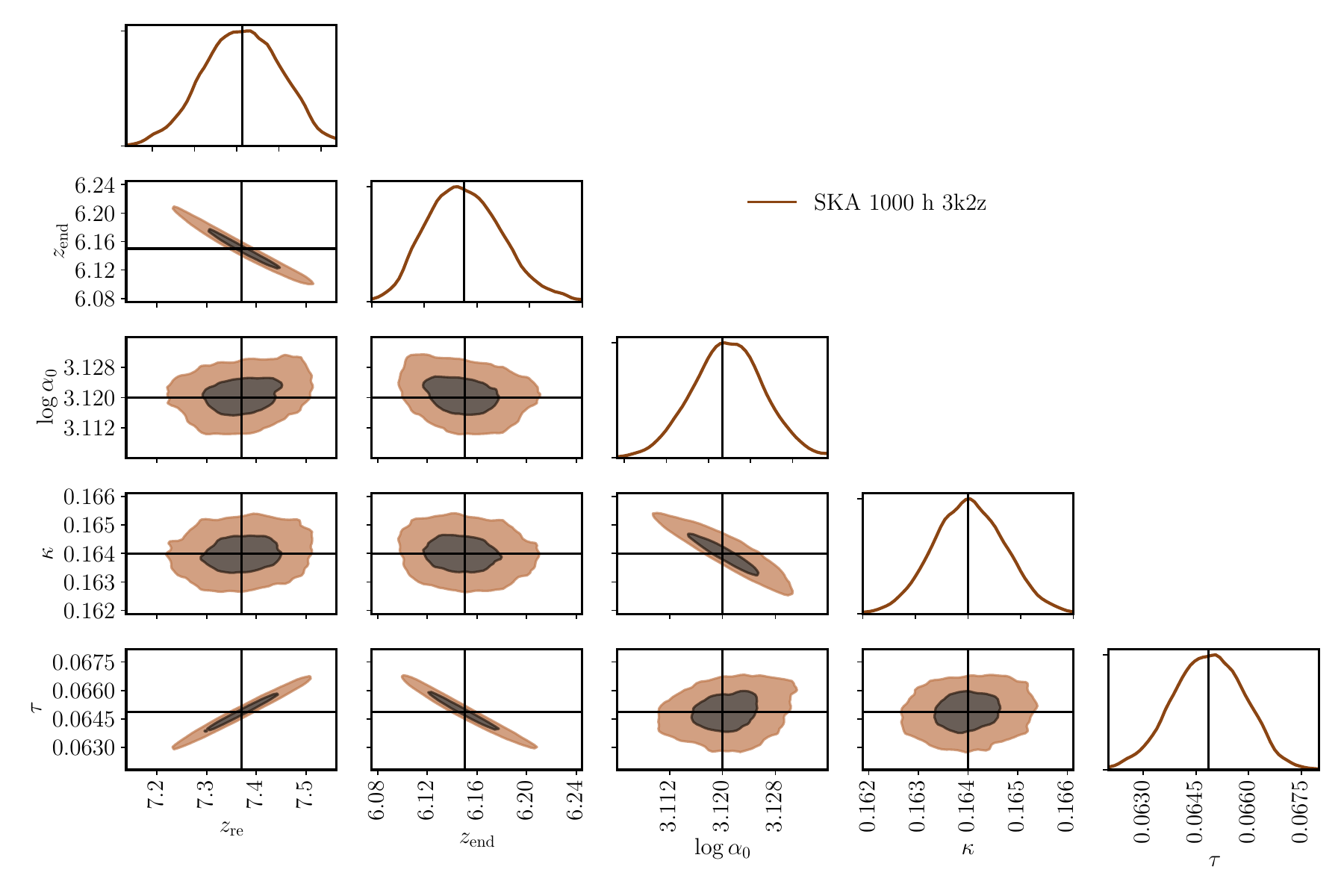}
    \caption{One- and two-dimensional posterior probability distributions for our model parameters and one derived parameter (the optical depth $\tau$) when fitting a detection of the pKSZ power spectrum measurement at $\ell = 3000$ with a ten per cent uncertainty and of the 21\,cm power spectrum at redshifts of $z =6.5, 6.5, 7.8$ at $k = 0.1, 0.50, 0.5$\,Mpc$^{-1}$ for 1000~hours of integration with SKA-\textit{Low}. For each parameter, the thin black vertical line is the true value used to generate the mock data and the dashed blue vertical line is the median of the distribution.}
    \label{fig::SKA_3k2z_triangle}
\end{figure*}

In this section, we investigate the possible degeneracies between our model parameters. We choose the \texttt{3k2z} model ($z = 6.5,6.5,7.8$ at $k = 0.1, 0.5, 0.5$ Mpc$^{-1}$) outlined in Sec.~\ref{sec::3k2z} for which the Gelman-Rubin convergence criteria in Table~\ref{tab:otherMCMC_results} is met ($R-1 \approx 0.001$). The corner plot in Fig.~\ref{fig::SKA_3k2z_triangle} clearly illustrates this, showcasing the relation between the global history and morphology parameters as well as the derived Thomson optical depth $\tau$. The parameter space is well sampled, with clear correlations seen only between either $z_{\mathrm{re}}-z_{\mathrm{end}}$ and log$_{10}(\alpha_{0})-\kappa$. A correlation is noticeable between the Thomson optical depth and each of the global history parameters. This is more than expected, as $\tau$ is an integral of the ionisation history, defined as a function of $z_\mathrm{re}$ and $z_\mathrm{end}$ in
equation~\eqref{eq::reion_parametrisation}. We discuss the constraints on $\tau$ for our different test cases in appendix.~\ref{appendix::tau} in more detail.

\section{Testing the forecast with {\sc RSAGE}}
\label{appendix::bias}

We have shown in Sec.~\ref{subsec:3_fit} that with as few as three data points (one kSZ, two 21\,cm), one can recover the global history of reionization and place some lower limits on the reionization morphology parameters. However, in these cases, both the mock data points and the sampled models were generated using our simplified derivation of the 21\,cm power spectrum, ignoring third-order terms and cross-correlations, given in equation~\eqref{eq:recons_21cm_ps_simplified}. In reality, of course, the data will include these extra terms. In this section, we will explore how this discrepancy can impact our forecast results.

We generate two new 21\,cm data points corresponding to the \texttt{1k2z} case of Sec.~\ref{subsec:3_fit}, for 100\,hours of observations with SKA-\textit{Low}. This time, the points are generated using the full expression of $P_{21}(k,z)$, including the cross- and higher-order terms, given in equation~\eqref{eq:recons_21cm_ps_full}. As discussed in Sec.~\ref{subsec:3_recons}, these points have an amplitude between 10 to 25\% smaller than what was obtained with the simplified expression. 

\begin{figure}
    \centering
    \includegraphics[width=\columnwidth]{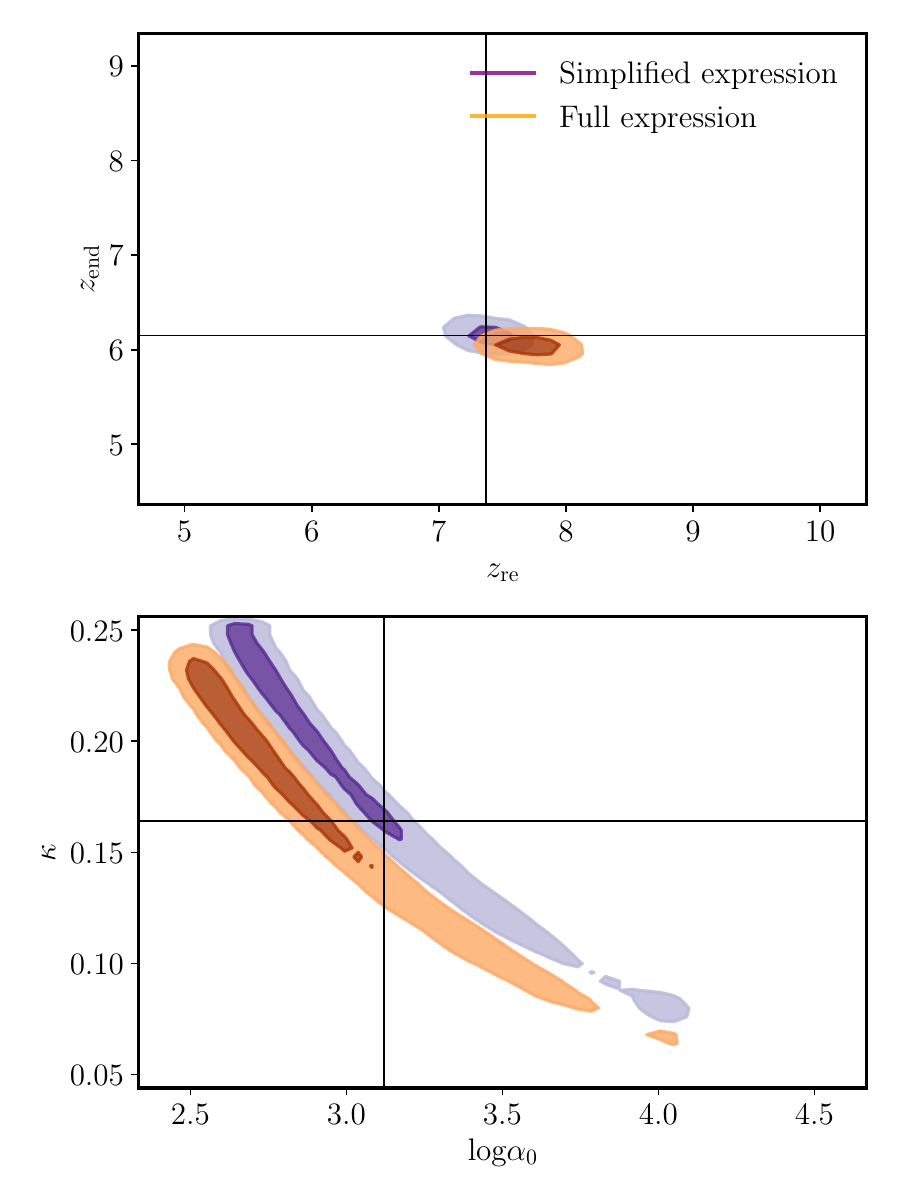}
    \caption{Posterior distributions on reionization global (\textit{upper panel}) and morphology (\textit{lower panel}) parameters when assuming two measurements of the 21\,cm power spectrum at $z=6.5$, 7.8 and $k=0.5\,\mathrm{Mpc}^{-1}$ for 100\, hours of observation with SKA-\textit{Low}. The mock 21\,cm data points used for the fit are generated either with the simplified equation~\eqref{eq:recons_21cm_ps_simplified}, in purple, or with the full equation~\eqref{eq:recons_21cm_ps_full}, in orange. Vertical and horizontal black lines correspond to the `true' values of the parameters, used to generate the mock data.}
    \label{fig:ideal_vs_rsage}
\end{figure}

We fit our four reionization parameters to these new points (note that the kSZ data points remain unchanged) through MCMC sampling and show the resulting two-dimensional posterior distributions in Fig.~\ref{fig:ideal_vs_rsage}. As expected from our discussions in Sec.~\ref{subsec:3_recons}, the general shape of the joint distributions is unchanged, only the results are now biased. As seen in Fig.~\ref{fig:recons_comp}, the simplified model tends to overestimate the 21\,cm power for a given redshift, having an amplitude equivalent to a lower ionisation level. For this reason, fitting it to unbiased data leads to an underestimate of the endpoint of reionization by $\Delta z_\mathrm{end}=0.12$. The midpoint is even more impacted as it is now biased by $\Delta z_\mathrm{re}=0.36$, nine times more than with the simplified mock data. Note that these modified reionization histories result in a large optical depth: $\tau = 0.070 \pm 0.002$. Hence, a way to partially mitigate these biases could be to impose a Gaussian prior on $\tau$.
Similarly, the posterior distributions of the morphology parameters are shifted compared to the results with the simplified mock data points.

Although these results exhibit a strong bias in all parameters, this bias is nevertheless well understood and easy to model. A potential way of improvement would be to add a nuisance parameter to the fit, as a pre-factor to the model 21\,cm power spectrum. This extra parameter would be marginalised over and would account for the extra power produced by the approximations done in equation~\eqref{eq:recons_21cm_ps_simplified}. Another option would be to precisely quantify this bias and systematically subtract it from the models, allowing one to reproduce the results of Sec.~\ref{subsec:3_fit}. However, this idea would be less successful when applied to real data, as this bias is model-dependent: From one simulation to another, the amplitude and shape of the cross- and higher-order terms missing in equation~\eqref{eq:recons_21cm_ps_simplified} will vary \citep{LidzZahn_2007, GeorgievMellema_2021}.

\section{Forecast constraints on the Thomson Optical Depth}
\label{appendix::tau}

\begin{figure}
    \centering
    \includegraphics[width=\columnwidth]{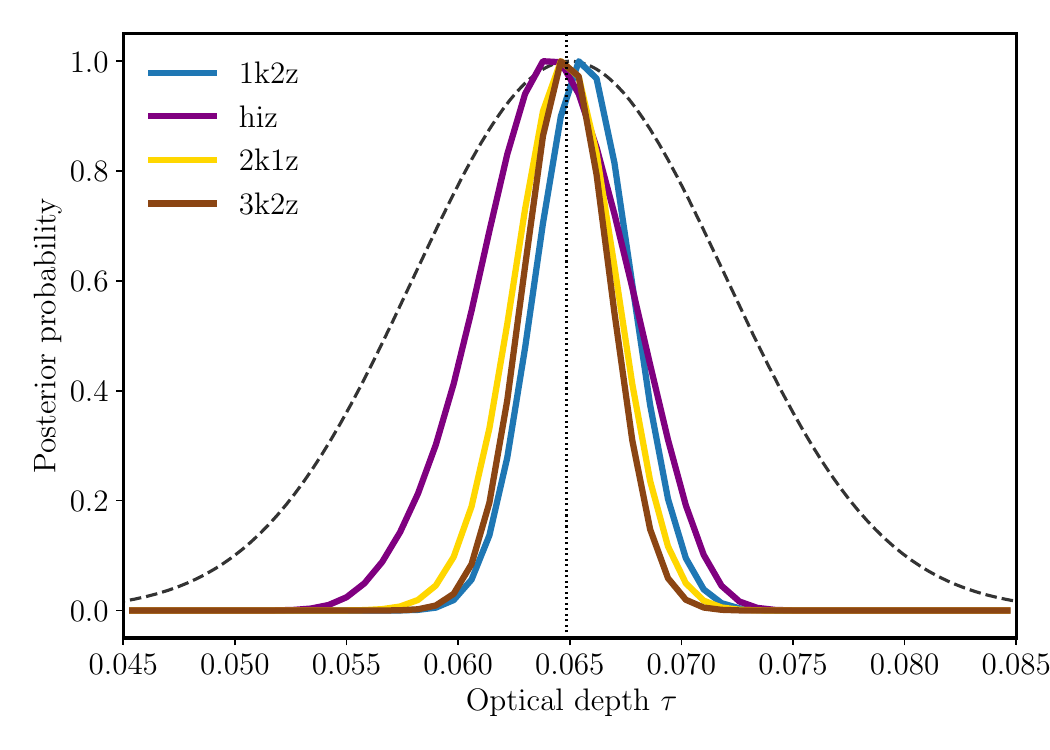}
    \caption{Posterior distributions of the Thomson optical depth for the range of models discussed within this work (see Tables~\ref{tab:MWA_MCMC_results} and \ref{tab:otherMCMC_results}). The dashed vertical line represents the `true' value $\tau = 0.065 $.}
    \label{fig:tau}
\end{figure}

In section.~\ref{sec::electron_powe_spectrum}, we derived the reionization history $x_v(z)$ using the global history parameters $z_{\mathrm{end}}$ and $z_{\mathrm{re}}$ (see equation~\ref{eq::reion_parametrisation}) and can, therefore, derive the Thomson optical depth $\tau$. With this in mind, we can investigate the constraining power of the forecast on $\tau$ from the combined 21\,cm and pkSZ power spectra analysis. Figure~\ref{fig:tau} presents the constraint on $\tau$ for each of the cases discussed throughout this work. The dashed line represents a Gaussian centred on the true model value $\tau = 0.0649$ (Table~\ref{tab:rsage_ref_params}) and with standard deviation the uncertainty reported in \citet{PlanckCollaborationAghanim_2020}. First, we note that, for each mock data set, we are able to recover the true value of the optical depth (values are biased by less than 1\%) while achieving a tighter constraining power than with current large-scale CMB data. Despite this, small variations in constraining power are visible between cases. 

Specifically, for our fiducial \texttt{1k2z} case (see Sec.~\ref{sec::SKA_ideal}), there is a visible positive bias, deviating from the true value by a fraction of a per cent. The bias arises from the choice of a fixed $k$-scale for the
21\,cm mock data points. Because reionization is a time-dependent and inhomogeneous process, the 21\,cm spectrum from the EoR is expected to evolve both with redshift and with $k$-scale. There is a characteristic $k$-scale below which the scale-dependence of $P_{21}$ is governed by the matter power spectrum. This characteristic scale evolves as reionization progresses \citep{Furlanetto2006}. Hence, $k$-scales lower than this characteristic scale will be less sensitive to reionization, and vice versa \citep[see fig.~5 and 6][]{GeorgievMellema_2021}. While mock data in the latter half of reionization ($z = 6.5, 7.8$) exclude larger values of $\tau$, data limited to $k=0.5$ Mpc$^{-1}$ slightly favour earlier reionization scenarios. The \texttt{hiz} case (in purple, $k=0.5$ Mpc$^{-1}$ for $z = 7.8, 10.4$) is biased towards lower values of the Thomson optical depth for similar reasons. The forecast is conducted on data at higher redshift, in the very early stages of reionization. The result is then a per cent bias of the derived optical depth and increased uncertainty, preferring a lower value of $\tau$ synonymous with a later reionization. It is only when the case where both 21\,cm mock data is at a fixed $z=6.5$ but at both high and low scales $k=0.1, 0.5$ Mpc$^{-1}$ (\texttt{2k1z} in yellow) are considered, that we can recover an unbiased estimate. This is because for the \texttt{2k1z} mock data, the progress of reionization is encoded differently in each of the $k$-scale of the 21\,cm power spectrum (see Sec.~\ref{subsec::2k1z}), resulting in a tighter constraint on $\tau$. Naturally, the \texttt{3k2z} case (seen in yellow), which combines the \texttt{1k2z} and \texttt{2k1z} mock data, can further constrain $\tau$ without bias and with a lower uncertainty. 
%In summary, the choice of redshift dependence of the 21\,cm mock data is what allows us to constrain the Thomson optical depth, however, choosing a fixed $k$-scale of the measurements will slightly bias the result.

%%%%%%%%%%%%%%%%%%%%%%%%%%%%%%%%%%%%%%%%%%%%%%%%%%

%%%%%%%%%%%%%%%%% APPENDICES %%%%%%%%%%%%%%%%%%%%%

%%%%%%%%%%%%%%%%%%%%%%%%%%%%%%%%%%%%%%%%%%%%%%%%%%

% Don't change these lines
\bsp	% typesetting comment
\label{lastpage}
\end{document}